\documentclass[aps,pra,superscriptaddress,groupedaddress,bibnotes,twocolumn,reprint]{revtex4-2}
\usepackage{graphicx}
\usepackage{comment}
\usepackage{dcolumn}
\usepackage{bm}
\usepackage{amsmath}
\usepackage{amssymb}
\usepackage{physics}
\usepackage{mathtools}
\usepackage{natbib}
\usepackage{hyperref}
\usepackage{xcolor}
\usepackage[utf8]{inputenc}
\usepackage[T1]{fontenc}
\usepackage{newtx}
\usepackage{caption}
\usepackage[labelfont={small}]{subcaption}
\usepackage{tikz}
\tikzset{meter/.append style={draw, inner sep=7, rectangle, font=\vphantom{A}, minimum width=20, line width=.5,
 path picture={\draw[black] ([shift={(.1,.3)}]path picture bounding box.south west) to[bend left=50] ([shift={(-.1,.3)}]path picture bounding box.south east);\draw[black,-latex] ([shift={(0,0.15)}]path picture bounding box.south) -- ([shift={(.25,-0.1)}]path picture bounding box.north);}}}
\usetikzlibrary{decorations.pathreplacing}
\usetikzlibrary{decorations.pathmorphing}
\usetikzlibrary{shapes.geometric}
\begin{document}

\title{
Trade-off between complexity and energy in quantum phase estimation
}

\author{Yukuan Tao}
\email{yukuan.tao@nottingham.ac.uk}
\affiliation{
 School of Mathematical Sciences and Centre for the Mathematics and Theoretical Physics of Quantum Non-Equilibrium Systems, University of Nottingham, University Park, Nottingham NG7 2RD, UK
}

\author{M\u{a}d\u{a}lin Gu\c{t}\u{a}}
\email{madalin.guta@nottingham.ac.uk}
\affiliation{
 School of Mathematical Sciences and Centre for the Mathematics and Theoretical Physics of Quantum Non-Equilibrium Systems, University of Nottingham, University Park, Nottingham NG7 2RD, UK
}

\author{Gerardo Adesso}
\email{gerardo.adesso@nottingham.ac.uk}
\affiliation{
 School of Mathematical Sciences and Centre for the Mathematics and Theoretical Physics of Quantum Non-Equilibrium Systems, University of Nottingham, University Park, Nottingham NG7 2RD, UK
}

\begin{abstract}
{Quantifying the energetic cost of implementing quantum operations is essential for assessing the efficiency and scalability of quantum sensing and information-processing technologies. Here, we introduce a framework for analysing the interplay between complexity and energy cost of quantum processes}. In particular, we {apply our framework to} a sequential quantum phase estimation protocol, where a phase of physical significance is encoded in a quantum channel. The channel is applied to a probe state repeatedly until the probe is measured and the outcome leads to an estimate on the phase. We establish a trade-off relation between {the total energy cost of the protocol} and the number of times the channel is applied (complexity), while reaching a desired estimation precision. {A sweet spot is located where the two quantities are co-optimised.} The principles of our analysis can be adapted to {benchmark the energetic requirements} in other quantum protocols and devices. 
\end{abstract}
\maketitle

\section{Introduction}\label{sec:realintro}
Quantum protocols are known to outperform their classical counterparts for various tasks. This so-called {\it quantum advantage} relies on different characteristics which are available in the quantum realm, such as coherent superpositions \cite{CohRes} and entanglement of quantum states \cite{EntRes,Chitambar_2016}. Identifying and characterising the {\it resources} enabling quantum advantages is a major goal of quantum information science, often studied under the umbrella of quantum resource theory \cite{Chitambar_2019}. The advantage itself may manifest in a variety of forms and can be quantified by means of different figures of merit, depending on the task at hand. In particular, by quantum advantage in {\it complexity} we mean that, while reaching the same goal, the total number of elementary operations performed by a quantum protocol is smaller than that by a classical one. For examples, in quantum computation \cite{Nielsen_Chuang_2010,harrow2017quantum}, Grover's algorithm \cite{grover1996fastquantummechanicalalgorithm} achieves a quadratic speed-up for the searching problem, and, more remarkably, Shor's algorithm \cite{Shor_1997} is able to attain an exponential speed-up for integer factorisation. Both are relevant to important computational fields such as cryptography \cite{Mavroeidis_2018}. Exponential speed-up is also possible for Hamiltonian simulation \cite{Lloydsim, Low_2019} --- following from the idea of simulating physical systems through quantum computers as originally raised by Feynman \cite{Feynman}. Meanwhile, in quantum metrology, there exists a quadratic speed-up to reach the desired estimation precision, termed the Heisenberg limit~\cite{Giovannetti_2004,Giovannetti2006,Giovannetti_2011}. This can help improve performance in a wide range of applications, including gravitational sensing \cite{bongs2002highorderinertialphaseshifts}, biological imaging \cite{Taylor_2013} and timing \cite{Katori2011OpticalLC}. 

Despite their promises, the above quantum advantages typically only hold under a noise-free setting. In realistic experimental situations, the target system inevitably interacts with its surrounding, being the experimental device or the inaccessible environment, causing decoherence on its evolution \cite{Zurek2003}. The resulting noisy quantum operation deviates from the desired one and the {\it implementation error} generally leads to a larger complexity, sometimes even losing the quantum advantage: under commonly encountered noise models, Refs.~\cite{Salas_2007,Shapira2003, Shenvi2003} and Ref.~\cite{Chuang_1995} show that Grover's algorithm and Shor's algorithm, respectively, can only attain partial complexity advantage if the noise is weak enough, while Refs.~\cite{Demkowicz_Dobrza_ski_2012, Rafa2010, Sekatski_2017} indicate that the metrological advantage is reduced to a constant factor unless specific structures of the noise are assumed. 

In order to retain the ideal quantum complexity, various error mitigation techniques have been designed \cite{Demkowicz_Dobrza_ski_2012, Rafa2010, Sekatski_2017,Lidar_Brun_2013,Zhou_2018,RafalQEC,Campbell2024,QEC2024}, and the resulting Noisy Intermediate-Scale Quantum (NISQ) technologies \cite{Preskill_2018} represent the state-of-the-art progress made towards commercialisation. Generally, to reduce the implementation error we can increase the power of the coupled device, or exert external controls to detect and correct the noise. Either way, complexity reduction comes with extra {energy} cost and a competition takes place between a smaller number of operations and larger energy cost per operation. In Refs.~\cite{AlexiaQEI,AlexiaStack,Alexia2021} the {\em total energy cost} of quantum protocols is treated as another important quantity that one wants to optimise for both scientific interest and near term realisation. From this perspective, since the optimal complexity --- corresponding to zero implementation error --- can be expensive, some finite error may indeed be preferred for an energy optimisation task. 

\begin{figure*}[thb]
\centering
\includegraphics[width=0.75\textwidth]{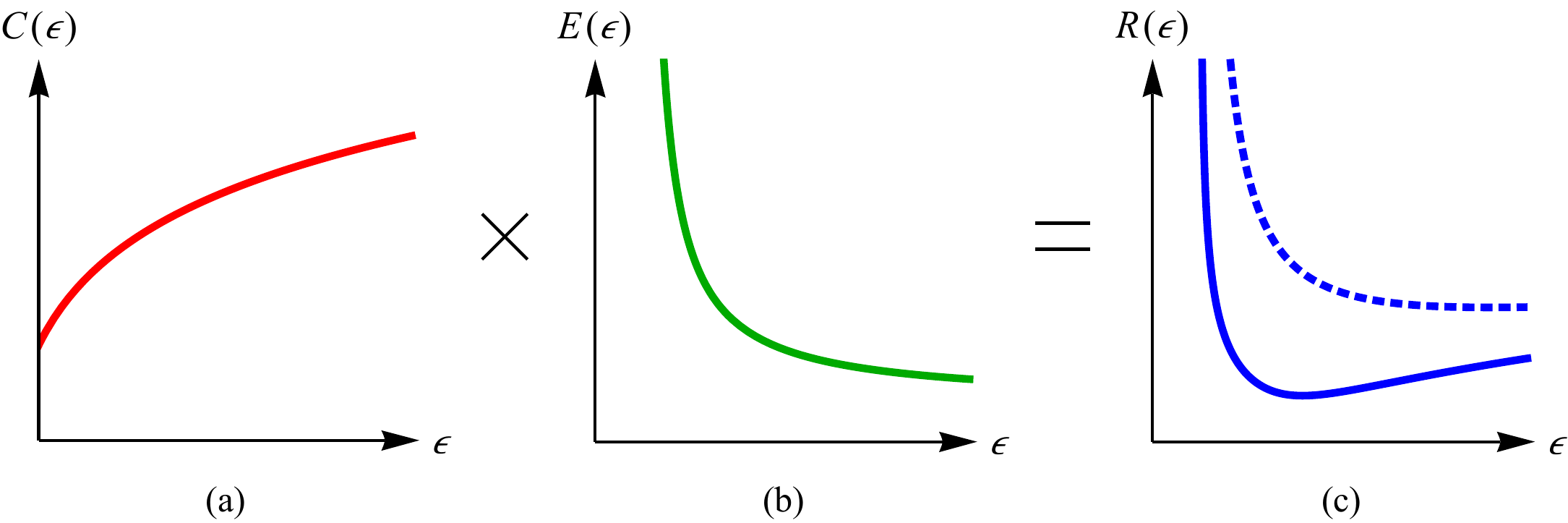}
\caption{A qualitative plot of the key variables characterising a quantum protocol as in Eq.~\eqref{noisyRes}. {Here $\epsilon$ is a generic quantifier for the implementation error of each elementary unitary gate, with $\epsilon \rightarrow 0$ corresponding to the ideal limit of perfect implementation.} Competition takes place between (a) growing gate complexity $C(\epsilon)$ and (b) decreasing energy cost $E(\epsilon)$ per gate as $\epsilon$ increases. (c) The resulting total energy cost $R(\epsilon)$ may (solid curve) or may not (dashed curve) have a global minimum at finite $\epsilon$.}\label{figure:resource}
\end{figure*}
Motivated by the quest {to explore more concretely the aforementioned trade-off between the key resources of gate complexity and energy cost}, we study the balance between the two in the context of {\em quantum metrology}. A typical metrological task consists of a pair $(\mathcal{G}_\phi,Strat)$, where $\mathcal{G}_\phi$ is an operation that encodes the parameter $\phi$ to be estimated and $Strat$ refers to strategies adopted to extract the parameter. The complexity here is represented by the number of times $\mathcal{G}_\phi$ is queried. A strategy consists of three main components: state preparation, intermediate controls (including error mitigation procedures) and measurements. Refs.~\cite{Lipka_Bartosik_2018,Liuzzo_Scorpo_2018,victor2025,chen2025optimalquantummetrologyenergy} have studied the cost of these components, while in most cases $\mathcal{G}_\phi$ is assumed to be the result of a free evolution and so does not incur additional cost. In this work we will introduce scenarios where energy is required to initiate the evolution. The form of $\mathcal{G}_\phi$ then depends on its implementation energy, and the larger this cost, the smaller the number of iterations of $\mathcal{G}_\phi$ is required to reach a desired estimation precision, leading to a {\em complexity-energy trade-off relation}. The corresponding implementation error is quantified by a distance between $\mathcal{G}_\phi$ and its ideal form. The main finding of this work exposes a critical error level for complexity-energy co-optimisation, beyond which the saving in one factor is overwhelmed by the overhead in the other.  We also expect that the combination of our results and the established complementary ones can lead to a more complete energy benchmarking in quantum sensing and metrology protocols.

The overall flow of the paper is as follows: Section~\ref{section:introduction} introduces the key concepts behind the trade-off relation that are expected to apply qualitatively to all relevant protocols, Section~\ref{ResQPE} describes the basics of the quantum phase estimation protocol in  metrology and its connection with the trade-off framework, while Section~\ref{section:firstprinciple} dives deeper into this connection by quantitatively analysing {a full sequential phase estimation protocol carried out on an optical platform, and drawing general insights on the interplay between complexity and energy cost in quantum processes}. The paper is concluded by a {summary and outlook} in Section~\ref{section:outlook}.

\section{The Trade-off Framework}\label{section:introduction}
Let us begin with a qualitative description of the origin of the trade-off. Suppose we have a fixed objective and a protocol to reach it. The protocol {takes energy to be implemented} and involves a certain notion of complexity. Both can depend on various factors, with the implementation {quality} being the focus of this work. In general, the {better} the protocol is implemented, the objective can be achieved with a smaller complexity, but {at the expense of a larger energy cost}. The optimal {quality} level can be determined accordingly.

To formalise the notion of complexity clearly, we focus on quantum tasks with the following ideal structure:
\begin{equation}\label{sequential}
    \rho_0 \xrightarrow{U^N} \rho_N\quad(\times\ Q_N).
\end{equation}
In words, an input state $\rho_0$ undergoes a sequence of $N$ identical unitary transformations $U$. Desired information accumulates in the final state $\rho_N$ and is often extracted through classical post-processing of measurement results. The whole sequence is repeated independently for $Q_N$ rounds in order to gain a set amount of total information. This set goal can be represented by a fixed constraint, denoted as $Con$, whose exact form depends on the task. The total number of gates implemented is $Q_N\cdot N$ and optimising over $N$ with respect to the fixed goal gives the (gate) {\em complexity} $C$ of the protocol:
\begin{equation}\label{def:complexity}
 C:=\min_N (Q_N\cdot N)|_{Con}.
\end{equation}

{In terms of the energy cost, the implementation of each gate is assumed to be independent and identically distributed (iid); therefore, we can denote $E$ as the {\em energy cost} for constructing each $U$-block. The {\em total energy cost}, denoted as $R$, is then 
\begin{equation*}
R= C\times E.
\end{equation*}}
However, perfect implementation can be affected by both experimental and fundamental limitations. The actual implemented gates will not be unitary in general and an error dependence needs to be added to the quantities defined so far:
\begin{equation}\label{noisyRes}
R(\epsilon)=C(\epsilon)\times E(\epsilon),
\end{equation}
where $\epsilon$ is some error parameter quantifying deviation from the desired unitary gate. A larger error {(i.e., lower implementation quality)} tends to increase the complexity{~\cite{Salas_2007,Shapira2003, Shenvi2003,Chuang_1995,Demkowicz_Dobrza_ski_2012, Rafa2010, Sekatski_2017, aharonov1999,gottesman2014,Lanzagorta2012ErrorScaling} }, as intuitively less information is encoded in each implemented gate, while reducing the energy cost per gate~\cite{Banacloche2002,Chiribella2021}.

{We are interested in scenarios where such a trade-off leads to an initially decreasing total energy cost $R(\epsilon)$ as a function of the error $\epsilon$}, as illustrated in Figure~\ref{figure:resource}. {Note that if $\epsilon$ --- and hence $E(\epsilon)$ --- is fixed, the total energy cost becomes equivalent to the gate complexity as a resource quantifier. It is thus precisely through the dynamic interplay between the two, as mediated by a variable implementation error, that we can introduce a novel resource analysis on the quantum protocol at hand.} In particular, minimising the complexity is not equivalent to minimising the total {energy cost $R(\epsilon)$} with respect to $\epsilon$. 
We also emphasise that the objective here is on optimising the quantum protocol itself, rather than determining if a quantum over classical energetic advantage exists; for works on the latter, we refer to Refs.~\cite{Meier_2025,Jaschke_2023,thompson2025energeticadvantagesquantumagents}. 

For a similar purpose to ours (see also the earlier analysis in Ref.~\cite{Liuzzo_Scorpo_2018}), Ref.~\cite{AlexiaQEI} brings up the necessity to build a framework within which costs of quantum protocols can be analysed and compared in a universal fashion. The proposed framework is dubbed Metric-Noise-Resource (MNR) \cite{AlexiaStack}: a metric is chosen to assess the performance, the effect of noise is taken into account, and the total resource cost is evaluated. A notion of efficiency is correspondingly defined as 
\begin{equation}\label{MNREff}
    \text{Efficiency } \eta=\frac{\text{Metric}}{\text{Resource}}.
\end{equation}

{For us, the constraint can be quantified by a suitably chosen metric, on which both resource cost (taken to be energy here) and complexity depend. Then for a fixed constraint, or equivalently a target value of the metric, we can minimise the total energy cost over $\epsilon$ to determine the optimal energy efficiency $\eta$.} We shall return to this connection in Section~\ref{sec:discussion}, after the energy analysis is carried out for a quantum phase estimation protocol performed on an optical platform, with the quantum Fisher information acting as the metric. 

{Finally, note that the iid assumptions underlying Eq.~\eqref{sequential} may be relaxed.  Quantum circuits used to perform quantum information processing tasks typically consist of unitary gates selected from a restricted set $\{U_1,...,U_n\}$ --- for example, composition of single-qubit Pauli gates and CNOT gates leads to universal quantum computation~\cite{Barenco_1995,Nielsen_Chuang_2010}, while in quantum simulation tasks one only has access to a limited set of control Hamiltonians~\cite{Lloydsim,Dalessandro2007}.
Each elementary gate can then be characterised by its own implementation error $\epsilon_i$, gate complexity $C_i(\epsilon_i)$, and energy cost $E_i(\epsilon_i)$, and the quantities in Eq.~\eqref{noisyRes} are extended to 
\begin{equation}\label{eqn:ResourceGeneral}
C=\sum_{i=1}^nC_i(\epsilon_i);\quad R=\sum_{i=1}^{n} C_i(\epsilon_i)\times E_i(\epsilon_i).
\end{equation}
Optimisation on the two types of resource is now over a set of error parameters for each elementary gate. Furthermore, one can introduce implementation schemes where energy is shared or recycled across multiple gates, resulting in spatiotemporal correlations among them. There is no more simple product form for computing the total energy cost, and we may only define an effective mean energy cost of implementing each gate $\bar E=R/C$. These generalisations would cover a wider range of practical schemes; nevertheless, our simpler yet  illustrative iid case will turn out to provide already valuable insight on the desired complexity-energy trade-off relation.}

\section{Quantum Phase Estimation}\label{ResQPE}
\subsection{Quantum Fisher Information and Complexity}\label{basic}
A quantum metrology task can often be recast in the form of quantum phase estimation (QPE). In general, a phase $\phi$ is encoded in some quantum operations which are applied to a probe state and by measuring the probe we can obtain an estimate on the phase. The precision of the estimation can be quantified by the quantum Fisher information (QFI), which captures geometrically the rate of change of the probe state under an infinitesimal variation of the phase parameter \cite{Caves1994}. To compute the QFI, we first introduce the classical Fisher information (CFI): for a positive operator-valued measurement (POVM) defined by the set of operators  $\textbf{M}=\{M_i\}$ and a probe state $\rho_\phi$ parametrised by the phase $\phi$, the CFI is
\begin{equation}\label{CFI}
\begin{aligned}
    &F_c[\textbf{M}](\phi)=\sum_i \Tr (\rho_\phi M_i)l_i^2, \\
    &l_i=\partial_\phi\log\Tr(\rho_\phi M_i)=\frac{\Tr(\dot{\rho}_\phi M_i)}{\Tr(\rho_\phi M_i)},
\end{aligned}
\end{equation}
where the dot denotes the derivative over $\phi$. 

The QFI is achieved by optimising the CFI over all possible POVMs:
\begin{equation*}
F_q(\phi)=\max_{\textbf{M}}F_c[\textbf{M}](\phi).
\end{equation*}
It turns out the optimal $\textbf{M}$ are projectors onto the eigenspaces of the symmetric logarithmic derivative (SLD) operator $\Lambda_\phi$, which satisfies the equation $\partial_\phi\rho_\phi=\frac{1}{2}(\Lambda_\phi\rho_\phi+\rho_\phi\Lambda_\phi)$. Suppose the same procedure is repeated independently for $Q$ times, by the additive nature of the QFI for product states \cite{T_th_2014}, the total QFI becomes $QF_q(\phi)$, and the quantum Cram\'er--Rao bound~\cite{holevo2011probabilistic} states that
\begin{equation}\label{cramer}
    \text{Var}[\hat{\phi}]\geq \frac{1}{QF_q(\phi)},
\end{equation}
where $\hat{\phi}$ is an unbiased estimator of $\phi$ based on the $Q$ measurement outcomes and $\text{Var}[\hat{\phi}]$ is its variance. Note that the bound is only achievable in the asymptotic limit of large $Q\gg1$ by employing adaptive estimation 
procedures for determining the optimal measurement 
\cite{Fujiwara2006,paris2009quantumestimationquantumtechnology}.

Within this framework we study the sequential strategy of the type described in Eq.~\eqref{sequential} as inspired by Refs.~\cite{Higgins_2007,Nichols2016,RafalSeq}: the phase $\phi$ is ideally imprinted via a unitary phase shift operator $\exp(-i\phi H{/2})$ where $H$ is a control Hamiltonian. The operator functions as an oracle and each time we make a query it is repeatedly applied to the probe state, which will carry accumulating information about the phase. Finally the state is measured, yielding an estimate on the phase. If the oracle is applied $N$ times, we denote the QFI of the final state by $F_N$. {In the rest of this work we will treat $N\geq1$ as a continuous variable for simplicity~\cite{RafalSeq,Dorner2009}.}

To determine the complexity, we quantify $Con$ in Eq.~\eqref{def:complexity} by demanding a target value, denoted as $\delta^2$, of the quantity on the right hand side of \eqref{cramer}, so that $\delta^2$ lower bounds the estimator variance. Naively, the number of independent repetitions of an $N$-step sequence needed to reach  the set goal is
\begin{equation}\label{numberrep}
q_N=\frac{1}{\delta^2 \cdot F_N},
\end{equation}
and so the complexity can be derived as
\begin{eqnarray}\label{rawcomplexity}
c&=&\min_N\big(q_N\cdot N\big)\big|_{\delta^2} \nonumber \\
&=&\frac{1}{\delta^2}\cdot\min_N \frac{N}{F_N} \\
&=& \frac{1}{\delta^2}\cdot \frac{N_{\textrm{opt}}}{F_{N_{\textrm{opt}}}}\,, \nonumber
\end{eqnarray}
where $N_{\textrm{opt}}$ is the optimal step for the minimisation and $|_{\delta^2}$ is a shorthand for the constraint $|_{[QF_q(\phi)]^{-1}\leq\delta^2}$. 

However, Eq.~\eqref{rawcomplexity} must be taken with a pinch of salt, and $c$ will be referred to as the \textit{raw complexity}{; the discrepancy arises due to the fact that $q_N$ as calculated by Eq.~\eqref{numberrep} is not necessarily an integer and hence implicitly assumes $F_N$ to be linear over $N$, while $Q$ in~\eqref{cramer} is an integer. In Appendix~\ref{app:rawvstrue} we elaborate on the general behaviour of $F_N$ and how its nonlinearity calls for a correction to $c$. To address this discrepancy, the \textit{true complexity} of the QPE protocol will be determined as follows. 

\begin{figure*}[th]
 \begin{subfigure}{0.49\textwidth}
     \includegraphics[width=\textwidth]{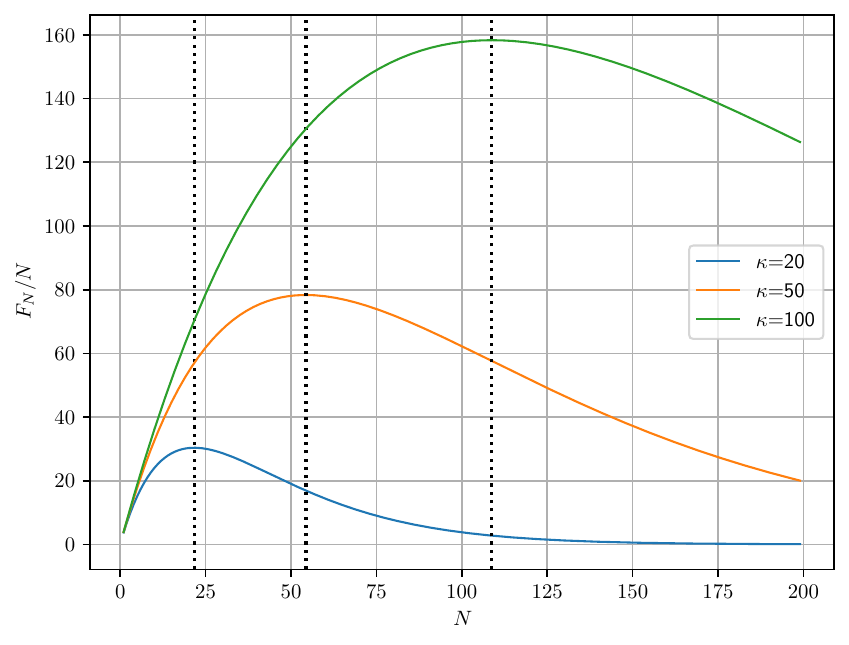}
     \subcaption{}\label{plot:QPEQFI}
 \end{subfigure}
 \hfill
 \begin{subfigure}{0.49\textwidth}
     \includegraphics[width=\textwidth]{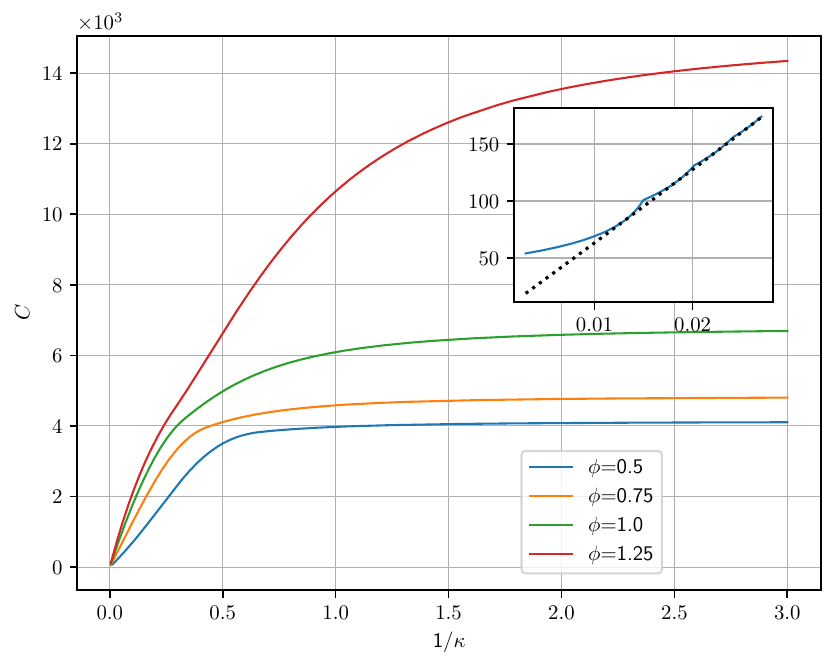}
     \subcaption{}\label{plot:QPEcomplexity}
 \end{subfigure}
\caption{(a) Plots of {the ratio between the QFI and the steps number}, $F_N/N$, for different values of {the concentration parameter of the von Mises-Fisher distribution} $\kappa$ and a fixed phase $\phi=0.5$ to be estimated. Observe the non-monotonic behaviour predicted by Eq.~\eqref{QFIapproxNichol}: the ratio grows linearly with $N$ first, reaches its maximum and decays exponentially afterwards. As indicated by the vertical dotted lines, the optimal step  $N_{\textrm{opt}}$ is well approximated by $-[2\log(\lambda_\perp)]^{-1}$; (b) The resulting complexity \eqref{QPEcomplexity} is plotted against $1/\kappa$ (representing the implementation error $\epsilon$) for {the desired lower bound on the estimator variance, $\delta^2=10^{-4}$, implied by Eq.~\eqref{cramer}}. Observe the growing pattern as anticipated by Figure~\ref{figure:resource}(a). The inset exhibits the zoom-in at small $1/\kappa$ for $\phi=0.25$. Each teeth of the zigzag pattern corresponds to a region that applies the same number of complete sequences $Q_N$. The dotted line corresponds to the raw complexity computed by Eq.~\eqref{rawcomplexity}. Notice that the raw complexity vanishes in the ideal limit, and approximates the true complexity better as $1/\kappa$ grows, {as discussed in Appendix~\ref{app:rawvstrue}.}\label{plot:QPEAB}}
\end{figure*}

Suppose an $N$-step sequence is first repeated fully for $Q_N=\lfloor q_N\rfloor$ times, $\lfloor q_N\rfloor$ being the integer part of $q_N$, followed by a final sequence with $N_0\leq N$ steps.} By additivity of the QFI, the overall QFI from the $(Q_N+1)$ sequences is $Q_NF_N+F_{N_0}$. {Thus, to meet the constraint set by $\delta^2$, $N_0$ must be such that
\begin{equation}\label{truelastroundstep}
F_{N_0}=\frac{1}{\delta^2}-Q_NF_N.
\end{equation}
The true complexity is then given by
\begin{equation}\label{QPEcomplexity}
C=\min_N\big(NQ_N+N_0\big)\big|_{\delta^2}.
\end{equation}
The minimisation will be computed numerically and the minimal point is anticipated to be near $N\approx N_{\textrm{opt}}$. As seen shortly and qualitatively explained in Appendix~\ref{app:rawvstrue}, the true complexity can be well approximated by the raw one when the implementation error is large enough. }

\subsection{Modelling the Gate Implementation Error}\label{sec:imperror}
To incorporate the implementation error, we assume that the (ideally unitary) quantum channels imprinting the phase in each round of the QPE protocol are in fact generated by a random Hamiltonian. For a qubit system, this amounts to taking $H_\textbf{n}=\textbf{n}\cdot\boldsymbol{\sigma}$, where $\textbf{n}$ is a unit Bloch vector sampled from some probability distribution $p(\textbf{n})$ and $\boldsymbol{\sigma}$ is the vector of Pauli operators. In the ideal limit, $p(\textbf{n})$ is a Dirac delta function at $\textbf{n}_0$, where $\textbf{n}_0$ is the Bloch vector of the desired control Hamiltonian $H$. Both the error and the QFI, $F_N\equiv F_N(p;\phi)$, depend on the distribution. We will use semicolon to separate free parameters that affect the implementation error. The resulting noisy gate (channel) is
\begin{equation}\label{QFIChannel}
\mathcal{G}_{p;\phi}(\cdot)=\int_{|\textbf{n}|=1}e^{-i\phi\textbf{n}\cdot\boldsymbol{\sigma}{/2}}(\cdot)e^{i\phi\textbf{n}\cdot\boldsymbol{\sigma}{/2}}p(\textbf{n})\text{d}\textbf{n}.
\end{equation}
Correspondingly, the channel performs random rotation on the Bloch sphere. Let $\textbf{s}$ be the Bloch vector of the probe qubit. It is transformed by the implemented gate as
\begin{equation}\label{QFIChannelBloch}
\textbf{s}\rightarrow\textbf{G}_{p;\phi}\,\textbf{s},\quad \textbf{G}_{p;\phi}=\int_{|\textbf{n}|=1}\textbf{R}_\textbf{n}(\phi)p(\textbf{n})\text {d}\textbf{n},
\end{equation}
where $\textbf{R}_\textbf{n}(\phi)$ represents the rotation around $\textbf{n}$ by an angle $\phi$.
\subsection{Computing Complexity: An Example}\label{complexityEx}
As an example of the model introduced in the previous two subsections, we first consider the following setup, studied by Ref.~\cite{Nichols2016}. The initial state is set to be $\rho_0=\ketbra{\psi_0}{\psi_0}$, $\ket{\psi_0}=(\ket{0}+\ket{1})/{\sqrt{2}}$ and the desired Hamiltonian is $H=\sigma_z$. The random Bloch vector is sampled from the von Mises-Fisher distribution \cite{Fisher1953DispersionOA}, $p_\kappa(\theta)= \frac{\kappa e^{\kappa \cos\theta}}{4\pi\sinh\kappa}$, where $\theta$ is the azimuthal angle, and $\kappa$ is the concentration parameter. $H_{\textbf{n}}$ becomes uniformly distributed as $\kappa\rightarrow0$, and is sharply peaked at $\sigma_z$ as $\kappa\rightarrow\infty$. This distribution can be seen as the counterpart of a Gaussian over the Bloch sphere and the implementation error can be represented by $\epsilon \sim 1/\kappa$. Furthermore, since $p_\kappa(\theta)$ has axial symmetry around $z$, the resulting quantum channel is phase-covariant (commuting with $\sigma_z$) and the transformation matrix \eqref{QFIChannelBloch} can be expressed in the form of \cite{Smirne2016}
\begin{equation}\label{covariant}
    \textbf{G}_{\kappa;\phi}=
    \begin{bmatrix}
        \lambda_\perp(\kappa;\phi)\cos \phi & -\lambda_\perp(\kappa;\phi)\sin \phi&0\\
        \lambda_\perp(\kappa;\phi)\sin \phi & \lambda_\perp(\kappa;\phi)\cos \phi &0\\
        0&0&\lambda_\parallel(\kappa;\phi)
    \end{bmatrix}.
\end{equation}

Using the formula for $F_N(\kappa;\phi)$ derived in Appendix C of Ref.~\cite{Nichols2016}, we compute the complexity \eqref{QPEcomplexity} and plot it in Figure~\ref{plot:QPEAB}(b). Note that the step that maximises the QFI is not $N_{\textrm{opt}}(\kappa;\phi)$ for the raw complexity \eqref{rawcomplexity}: the quantity to be maximised here is $F_N/N$ (plotted in Figure~\ref{plot:QPEAB}(a)) rather than $F_N$. The latter is shown to be well approximated by
\begin{equation}\label{QFIapproxNichol}
    F_N(\kappa;\phi)\approx N^2\lambda_\perp^{2N-2}\left[\lambda_\perp^2+(\partial_\phi\lambda_\perp)^2\right],
\end{equation}
so that $F_N$ and $F_N/N$  attain their maximum at around $-[\log(\lambda_\perp)]^{-1}$ and $N_{\textrm{opt}}=-[2\log(\lambda_\perp)]^{-1}$, respectively. 

With the a priori probability distribution comes the lack of knowledge on physical details on how the phase-encoding channel is achieved. Consequently, only an educated guess can be made on the energy cost of the implementation. In the next Section, we restart with another more practical example and approach the QPE protocol from its physical foundation to realise a comprehensive resource analysis.

\section{Construction from First Principles}\label{section:firstprinciple}
In this Section, we adopt the sequential procedure from Section~\ref{ResQPE} for {a physically motivated task based on light--matter interaction. 
 A coherent laser pulse is used to drive a probe qubit, and the unknown field-qubit 
coupling strength \(g\) appears as the rotation angle of the induced 
unitary gate. Repeated applications of this gate sequentially encode the parameter
\(g\) in the probe state, allowing the coupling strength to be 
estimated through the QPE protocol.}

{The microscopic construction enables both the implementation error and 
the associated energy cost to be derived from the same physical model, 
rather than postulated as in Section~\ref{complexityEx}. For each 
application of the phase-encoding gate, the field is initialised from the vacuum in a coherent state, then allowed to 
interact with the probe qubit, and subsequently discarded. 
The field is 
therefore not reused between successive gates, so no field memory is 
retained~\cite{Yang_2019}, resulting in a Markovian process. Accordingly, the energy cost of each gate is the initialisation energy required 
to prepare the corresponding laser pulse, evaluated as the increase in 
the field energy relative to its vacuum state.} The effect of the implementation error on the complexity is also derived. Combining both factors gives us the total {energy} cost of the estimation protocol. 
{For completeness} we also include the cost of state preparation and measurement that take place at the beginning and end of each sequence, respectively.
Figure~\ref{fig:circuitcool} at the end of this Section exhibits the overall circuit structure, while Table~\ref{tabula} below makes clear the meaning of each parameter to be defined in the following subsections. 

\begin{table}[th]
\begin{tabular}{ c|l } 
\hline\hline
\multicolumn{2}{c}{Variable Parameters} \\
 \hline
 $\bar{m}$ & \small{number of photons spent per gate}  \\ 

 $\delta^2 $ & \small{lower bound on the estimator variance as implied by \eqref{cramer}}\\ 

 $M_s$ & \small{number of cooling qubits per sequence during state preparation} \\ 

 $M_m$ & \small{number of cooling qubits per sequence during measurement}\\
 \hline
 \multicolumn{2}{c}{Fixed Parameters}\\
 \hline
 $g$ & \small{field-qubit coupling strength (the ``phase'' to be estimated)}\\

 $T_0$ & \small{temperature of the free qubits used for cooling}\\

 $\omega_0$ & \small{transition frequency of the system qubit} \\
 
$\omega_1$ & \small{transition frequency of the pointer qubit}\\

 $\omega$ &  \small{frequency of the electromagnetic field}\\
 \hline\hline
 \end{tabular}
 \caption{\label{tabula}Summary of parameters relevant to our model.}
 \end{table}


\subsection{{Energy Bookkeeping}}\label{sec:energynotion}
\begin{table*}[t!]
{
\begin{tabular}{ c|lll } 
 \hline \hline
Building block & $\rho$&$V$ & $O$\\
\hline
Implementation of $n^{\text{th}}$ gate& $\rho_{n-1}^{(S)}\otimes\rho_{\text{vac}}$&Displacement operator on the field & Field-qubit free evolution through the coupling~\eqref{jaynes}\\
&&&Discarding the post-evolution field state\\
 &&&\quad \\
 State preparation & $\rho_\beta$& Cooling unitary & Discarding the auxiliary qubits\\
 &&&\quad \\
 Measurement  & $\rho_N^{(S)}\otimes\ketbra{0}{0}^{(P)}$&CNOT between system and pointer qubits \quad& Measurement on the pointer qubit \\
 &&& Discarding the probe qubit\\
 \hline \hline
 \end{tabular}}
\caption{{Summary of the initial state $\rho$, the non-energy-conserving unitary operation $V$, and the subsequent energy-conserving operations $O$ in each building block of the QPE protocol, as illustrated in Figure~\ref{fig:energynotion}. Here $S$, $P$ label the system probe qubit and pointer qubit, respectively, $\rho_{\text{vac}}$ is the vacuum state of the EM field, and $\rho_\beta$ is the thermal state of the cooling auxiliary qubits, as introduced in the following subsections. }}\label{tableenergy}
 \end{table*}
{Before proceeding, we need to establish a consistent operational definition of energy cost, which will be applied throughout the remainder of the paper. 
Each building block of the QPE protocol can be described by an initial quantum state $\rho$ of the system under consideration, 
assumed to be provided at no cost, followed by a target unitary operation $V$, after which 
only energy-conserving operations $O$ are allowed (including unitary evolution, discarding 
subsystems, and measurements). The full QPE circuit is constructed through concatenation of such building blocks: the output quantum state of one block is treated as the free input state for the next one. 

We define the energy cost of a block as the change in the system energy induced by $V$,
\begin{equation}\label{def:energy}
E = \Tr\!\left[ H_0\,(V\rho V^\dag - \rho) \right],
\end{equation}
where $H_0$ denotes the system's Hamiltonian.  This definition follows directly from energy conservation \cite{Chiribella2021}: any change in the system 
energy must be supplied by an external energy storage device, referred to as a battery.  Accordingly, $E$ quantifies the energy exchanged with the battery per 
application of $V$.

Operationally, $V$ can be realised by coupling the system to a battery state $\rho_B$ 
via an energy-conserving joint unitary $\tilde V$ satisfying 
$[\tilde V, H_B \otimes I_0 + I_B \otimes H_0] = 0$ (where $H_B$ denotes the battery's Hamiltonian), followed by decoupling \cite{Banacloche2002,Karasawa_2009, Ozawa_2002,Tajima2018,Tajima2020,Chiribella2021, tajima2025}:
\begin{equation}
\Tr_B\!\left[\tilde V (\rho_B \otimes \rho)\tilde V^\dag\right] \approx V\rho V^\dag.
\end{equation}
Following \cite{Chiribella2021,chen2025optimalquantummetrologyenergy}, we implicitly assume access to a battery with sufficiently large energy storage to implement all $V$ operations across the protocol~\cite{giulionote}. 

\begin{figure}[h!]
    \centering
    \begin{tikzpicture}[scale=1.1]
     \draw (0.25,0.1) node{\large $\rho$};
     \draw (0.5,0.5)--(1,0.5);
     \draw (0.5,0.3)--(1,0.3);
     \filldraw (0.75,0.15) circle (0.5pt);
     \filldraw (0.75,0) circle (0.5pt);
     \filldraw (0.75,-0.15) circle (0.5pt);
     \draw (0.5,-0.3)--(1,-0.3);
     \draw (1,-0.5) rectangle (2,0.7)node[anchor=center,midway]{\large $V$};
     
    \draw (2,0.5)--(2.25,0.5);
     \draw (2,0.3)--(2.25,0.3);
     \filldraw (2.125,0.15) circle (0.5pt);
     \filldraw (2.125,0) circle (0.5pt);
     \filldraw (2.125,-0.15) circle (0.5pt);
     \draw (2,-0.3)--(2.25,-0.3);
     \draw (2.25,-0.5) rectangle (3.25,0.7)node[anchor=center,midway]{\large $O$};

     \draw(3.25,-0.3)--(3.75,-0.3);
     \filldraw (3.5,-0.23) circle (0.4pt);
     \filldraw (3.5,-0.16) circle (0.4pt);
     \filldraw (3.5,-0.09) circle (0.4pt);
     \draw(3.25,-0.02)--(3.75,-0.02);
\node[cylinder, draw, 
shape border rotate = 90,
minimum width = 0.35cm,
    minimum height = 0.45cm, aspect=0.5] (c) at (1.5,1.5) {};
     \draw [-stealth] (1.5,1.35) -- (1.5,0.72);
     \draw(1.75,1.035) node{\large $E$};
 \end{tikzpicture}
    \caption{{An elementary building block used to quantify the energy cost: a free initial state $\rho$ undergoes a non-energy-conserving unitary $V$, followed by energy-conserving operations collectively represented by $O$. The energy cost $E$ is counted as the energy change of the state caused by $V$, supplied by a battery. The quantum output leg of $O$ is treated as the initial state for the next elementary block.}}
    \label{fig:energynotion}
\end{figure}

We shall adopt the above as a bookkeeping convention throughout: different components of the 
protocol correspond to different choices of $\rho$ and $V$, but the defining principle of $E$ 
remains unchanged. Figure~\ref{fig:energynotion} illustrates this energy accounting, and 
Table~\ref{tableenergy} summarises the corresponding objects for each building block of 
the QPE protocol. This convention isolates the fundamental implementation cost 
associated with non-energy-conserving operations, leaving aside additional 
thermodynamic or hardware overheads. These are briefly considered in Appendix~\ref{sec:Landauer} for the protocol at hand.}

\subsection{Gate Implementation}\label{EngtoComp}
Let the target system qubit (probe) be governed by the Hamiltonian $-\hbar\omega_0\sigma_z/2$, where $\omega_0$ is the transition frequency of the two-level system. The qubit is coupled to a resonant monochromatic electromagnetic (EM) field of frequency 
$\omega\,(\approx \omega_0)$ travelling in the $z$-direction. Suppose the rotating wave approximation \cite{RWA} holds, $|\omega_0-\omega|\ll\omega_0+\omega$. In the interaction picture, the system is then governed by a time-independent Hamiltonian \cite{Julio2002,JaynesCummings}:
\begin{equation}\label{jaynes}
H_S=\hbar gk_0(E\ketbra{1}{0}+E^*\ketbra{0}{1})=\hbar gk_0 E_0(\textbf{n}\cdot\boldsymbol{\sigma}),
\end{equation}
where $k_0$ is a unit quantity (time$^{-1}$) to keep the coupling constant $g$ dimensionless, $\textbf{n}=[\cos\theta,{\sin\theta},0]^\intercal$ and $E=E_0e^{i\theta}$ is the amplitude of the field with phase $\theta$. Physically, the parameter $g$ reflects the magnitude of the qubit dipole moment. To estimate it, we set the initial state to be $\rho_0=\ketbra{0}{0}$ and aim to implement the unitary gate 
\begin{equation*}
U_g=e^{-ig\sigma_x{/2}}. 
\end{equation*}
Classically, this is achieved by tuning $\theta=0$ and $2k_0E_0 t=1$, $t$ being the total evolution time. Quantum fluctuation of the field, however, sets a fundamental limit on the implementation quality: semi-classically, $E_0$ and $\theta$ can be treated as random variables with means $\bar{E_0}$ and $0$, respectively, and the evolution time is $t=1/(2k_0\bar{E_0})$. The gate implemented is then sampled from the set of unitary operators consisting of
\begin{equation}\label{unitarypulse}
    e^{-iH_St/\hbar}=\exp{-ig\frac{E_0}{\bar{E_0}}\textbf{n}\cdot\boldsymbol{\sigma}/2}=\exp{-ig\sqrt{\frac{m}{\bar{m}}}\textbf{n}\cdot\boldsymbol{\sigma}/2},
\end{equation}
where $m$ is the (coherent) photon number of the EM field, treated as a random variable with mean $\bar{m}$. The ideal limit corresponds to $\bar{m}\rightarrow\infty$ when $U_g$ is exactly applied to the qubit and the error may thus be characterised by $\epsilon\sim 1/\bar{m}$. For small $\bar{m}$, quantum statistics becomes significant and we have to resort to a full quantum treatment \cite{Igeta2013,Desutsch2004}, where random variables are further replaced with quantum operators. 
\begin{figure*}[th]
 \begin{subfigure}{0.46\textwidth}
     \includegraphics[width=\textwidth]{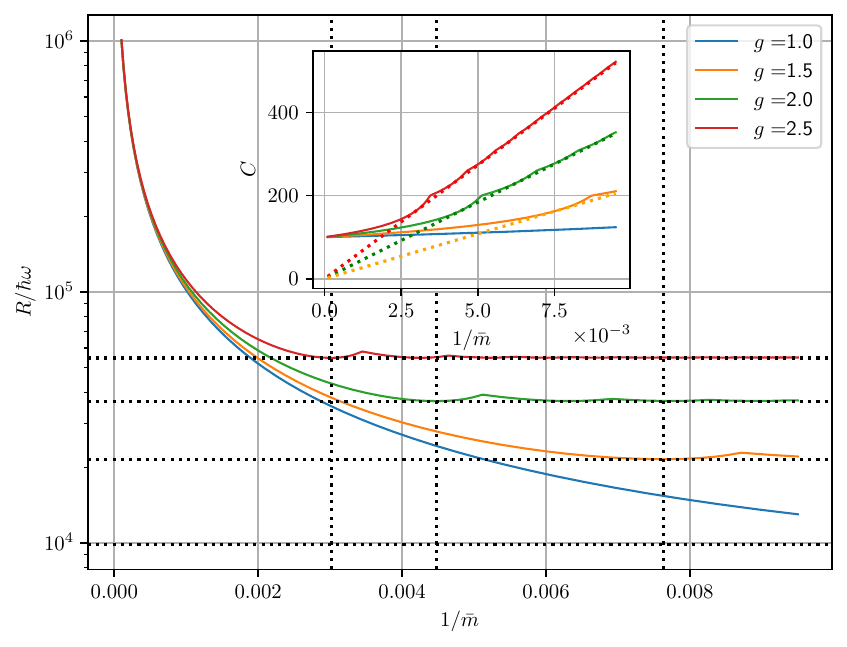}
     \caption{}\label{plot:complexityQPE}
 \end{subfigure}\hspace*{0.2cm}
 \begin{subfigure}{0.52\textwidth}
     \includegraphics[width=\textwidth]{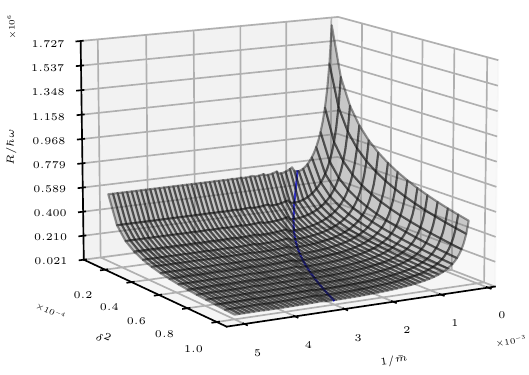}\caption{}
\end{subfigure}
\caption{(a) Plots of the total {energy} cost $R$~\eqref{ResourceQPE2}, {with the desired lower bound on the estimator variance set to  $\delta^2=10^{-4}$}, and the implementation error $\epsilon$ represented by $1/\bar{m}$, {i.e., the inverse of the number of photons spent on implementing each phase-encoding gate. As reflected by the $y$-axis label, $R$ is computed in units of the photonic energy of the EM field, $\hbar\omega$, and so numerically equals the total number of photons spent in implementing all the gates. The range of $\bar m$ is such that the condition after Eq.~\eqref{QFIapproximated} is met for our chosen values of the unknown coupling $g$, so that the QFI is well approximated by its leading order term.} 
The vertical and horizontal dotted lines locate the saturation point \eqref{optimalpoint} where the energy plateau starts. This matches the behaviour indicated by the dashed curve in Figure~\ref{figure:resource}(c). Solid and dotted lines in the inset are the corresponding true and raw complexities \eqref{approxQFIComp}, respectively. {As already observed in Figure~\ref{plot:QPEAB}(b) and predicted in Appendix~\ref{app:rawvstrue}, the latter approximates the former better as the error increases}. (b) The total energy plot in (a) is repeated for various $\delta^2$ and fixed $g=2.5$. The blue solid curve depicts the evolution of the saturation point as both $\bar{m}$ and $\delta^2$ vary.}\label{plot:QPEEnergy}
\end{figure*} 

Appendix~\ref{appendix:computingQFI} shows that in the semi-classical regime the QFI is well approximated by
\begin{equation}\label{QFIapproximated}
\begin{aligned}
&F_N\equiv F_N(\bar{m};g)\approx N^2 r^{2N},\\ &r\approx 1-\frac{\Delta(g)}{2\bar{m}},\quad  \Delta(g)=\frac{g^2+1-\cos(g)}{4},
\end{aligned}
\end{equation}
provided $\bar{m}\gtrsim100$, $g^2/\bar{m}\ll1 $ and the EM field is in its coherent state, representing lasers. Geometrically, the transformation on the $yz$ plane of the Bloch sphere by the implemented channel \eqref{QFIChannelBloch} is reduced approximately to a rotation of angle $g$ (corresponding to the desired unitary $U_g$) combined with a shrinking of factor $r$, matching the phase-covariant channel \eqref{covariant}. The optimal step for the raw complexity \eqref{rawcomplexity} and the corresponding number of repetitions \eqref{numberrep} are
\begin{equation}\label{optimalstepapprox}
\begin{aligned}
N_{\textrm{opt}}&\equiv N_{\textrm{opt}}(\bar{m};g)=-\frac{1}{2\log(r)}\approx -\frac{1}{2
\log\left(1-\frac{\Delta(g)}{2\bar{m}}\right)}\approx\frac{\bar{m}}{\Delta(g)};\\
q_{N_{\textrm{opt}}}&\equiv q_{N_{\textrm{opt}}}(\bar{m};g,\delta^2)=\frac{1}{\delta^2\cdot F_{N_{\textrm{opt}}}}\\ &\approx \frac{1}{\delta^2}\cdot \left(\frac{\Delta(g)}{\bar{m}}\right)^2\cdot \left(1-\frac{\Delta(g)}{2\bar{m}}\right)^{-\frac{2\bar{m}}{\Delta(g)}}\approx \frac{e}{\delta^2}\cdot\left(\frac{\Delta(g)}{\bar{m}}\right)^2.
\end{aligned}
\end{equation}
The raw complexity can be computed as
\begin{align}\label{approxQFIComp}
\begin{split}
c(\bar{m};g,\delta^2)&=q_{N_{\textrm{opt}}}\cdot N_{\textrm{opt}}\approx\frac{e}{\delta^2} \cdot\frac{\Delta(g)}{\bar{m}}.
\end{split}
\end{align}
This is a good approximation of the true complexity \eqref{QPEcomplexity}, $C(\bar{m};g,\delta^2)$, for large enough $1/\bar{m}$, as observed in the inset of Figure~\ref{plot:QPEEnergy}(a). 

{Moreover, we now hold knowledge on physical details of the implementation and hence its energy cost. To construct the $n^{\text{th}}$ phase-encoding channel in a sequence, we follow the established procedure in Figure~\ref{fig:energynotion}. As anticipated in Table~\ref{tableenergy}, the initial state  $\rho$ here is the tensor product of the probe state after the $(n-1)^{\text{th}}$ step $\rho_{n-1}^{(S)}$ and the vacuum state $\rho_{\text{vac}}$ of the EM field, the non-energy-conserving unitary $V$ is the displacement operator, $\mathcal{D}\left(\sqrt{\bar m}\right)$, on the field that creates the coherent state of mean photon number $\bar m$, and the ensuing energy-conserving operations $O$ consist of the free evolution through the field-qubit coupling~\eqref{jaynes} followed by a complete discarding of the field state at the end of the evolution. The output probe state $\rho_n^{(S)}$ then acts as part of the free initial state for the $(n+1)^{\text{th}}$ step, or as the pre-measurement state $\rho_N^{(S)}$ after the final step.} {More precisely, the quantities entering in the energy-defining Eq.~\eqref{def:energy} are
\begin{align*}
&\rho=\rho_{n-1}^{(S)}\otimes\rho_{\text{vac}},\quad
V=I_2\otimes\mathcal{D}\left(\sqrt{\bar m}\right),\\
&H_0=-\hbar\omega_0\frac{\sigma_z}{2}\otimes \mathcal{I}^{(F)}+\mathcal{I}^{(S)}\otimes\hbar\omega\hat n,
\end{align*}
where $\hat n$ is the number operator on the field, and $\mathcal{I}^{(S)},\mathcal{I}^{(F)}$ are the identity operators on the system qubit and the field, respectively. Since $V$ acts trivially
on the probe, the probe contribution to the energy change vanishes, while the energy cost of the displacement operator, as computed by Eq.~\eqref{def:energy}, is the energy difference between the vacuum and the coherent state, $E=\bar m\hbar\omega$, supplied by an external battery as in Fig.~\ref{fig:energynotion}.} 
For convenience, we will set the photonic energy to be unit, $\hbar\omega=1$, and so $E(\bar{m})=\bar{m}\sim1/\epsilon$. The total {energy} cost~\eqref{noisyRes}  is then
\begin{equation}\label{ResourceQPE2}
R\equiv R(\bar{m};g,\delta^2)=C(\bar{m};g,\delta^2)\cdot \bar{m}.
\end{equation}

{Note that here $E(\epsilon)$ diverges in the ideal limit of a perfect unitary implementation ($\epsilon\rightarrow0$ or $\bar m\rightarrow\infty$) and so the total energy cost $R(\epsilon)$ is expected to decrease initially, matching the qualitative pattern illustrated in Figure~\ref{figure:resource}. {We shall return to this point in Section~\ref{sec:discussion} where more general scenarios are considered.}

We plot $R$ in Figure~\ref{plot:QPEEnergy} within the approximation range. Starting from the ideal limit $1/\bar{m}\rightarrow0$, the total {energy} cost goes through a sharp drop, then plateaus after a saturation point. Since the complexity increases with $1/\bar{m}$, this point may serve as the {\em sweet spot} for complexity-energy  co-optimisation, beyond which energy saving becomes inefficient with respect to complexity overhead. Reading from the graph, the plateau appears around when more than one repetition of the sequence takes place and the sweet spot can be located by setting $q_{N_{\textrm{opt}}}=1$ in Eq.~\eqref{optimalstepapprox} and using Eqs.~\eqref{approxQFIComp}, \eqref{ResourceQPE2}, which leads to
\begin{equation}\label{optimalpoint}
\bar{m}_0\approx\Delta(g)\sqrt{\frac{e}{\delta^2}}\,,\quad C_0\approx \sqrt{\frac{e}{\delta^2}}\,,\quad R_0\approx\frac{e\Delta(g)}{\delta^2}\,.
\end{equation}
From Eq.~\eqref{approxQFIComp}, the product $c\cdot \bar{m}\approx R_0$ does not depend on $\bar{m}$, explaining the flattening up to the leading order. 

{Two remarks are in order. First, the implementation error of the phase-encoding channel, as modelled in this Section, comes solely from the quantum fluctuation of the EM field. In particular, we treat the field parameters as random variables instead of quantum operators and only extract the leading order effect to allow an analytical derivation. In doing so we have ignored other sources of noise, including laser inefficiency, decoherence caused by system-environment interaction, and higher-order effects from the field fluctuation which become significant if the field power is reduced. 
}

{Second, a caveat in identifying the sweet spot concerns the attainability of the quantum Cram\'er--Rao bound~\eqref{cramer}. 
Recall that the bound is tight only in the asymptotic limit, requiring in principle many repetitions, whereas the energy-optimal regime identified here involves only a small number of repetitions. 
Our analysis therefore uses the QFI metric as a benchmark rather than a guarantee of achievable performance. Finite-sample effects may shift the precise location of the optimum, but do not alter the qualitative trade-off between implementation quality and repetition cost. 
As shown in Appendix~\ref{app:variance}, operating at a smaller photon number per gate $\bar m$, in particular in the regime $10^2 \lesssim \bar m \lesssim 5 \cdot 10^2$, 
improves the consistency between the actual estimator performance and the asymptotic bound.}
\subsection{Other Sources of Cost}\label{section:othercost}
\subsubsection{Complexity and Energy Minimisation}\label{sec:optimalstepmod}
Apart from gate implementation, other components of the protocol also {cost energy to implement} and they may have impact on the complexity as well. We will next consider a few of these contributions for a more complete energy {bookkeeping (see also Appendix~\ref{sec:Landauer})}, namely the tasks of state preparation and measurement at the beginning and end of each sequence, respectively. Before proceeding, we first describe qualitatively the difference brought by the additional cost.

The two key quantities involved in the trade-off relation as illustrated in Figure~\ref{figure:resource} are the complexity $C$, or the minimal number of gates needed to reach a set objective~\eqref{QPEcomplexity}, and the corresponding total energy cost $R$. For the latter, 
we now take into account the additional (or `external') cost of each sequence other than the one spent on gate implementation, denoted as $E_{\textrm{ext}}$. Here it is treated as a constant for simplicity, while in general it may depend on parameters that can also affect the complexity, as we will see in Section~\ref{sec:stateprep}. The total {energy} cost as a function of the sequence steps number $N$ is accordingly modified as 
\begin{align}\label{totalcostext}
\begin{split}
&R_N\equiv R_N(\bar{m};g,\delta^2,E_{\textrm{ext}})\\
=&{
\begin{cases}
\underbrace{(NQ_N+N_0)}_{\text{number of gates}}\,\times\,\,\bar{m}\,\,\,+\underbrace{(Q_N+1)}_{\text{number of sequences}}\times \,\,E_{\textrm{ext}} & \text{if }N_0>0;\\ & \\
NQ_N\times\bar m+Q_N\times E_{\textrm{ext}} & \text{if }N_0=0.
\end{cases}}
\end{split}
\end{align}
We may denote $N_C$ and $N_R$ as the steps that minimise the total number of gates and the total {energy} cost, respectively, with the same constraint as in Eq.~\eqref{QPEcomplexity}:
\begin{align}\label{twooptimalsteps}
\begin{split}
 &N_C\equiv N_C(\bar{m};g,\delta^2)=  \arg\min_{N} \big(NQ_N+N_0\big)\big|_{\delta^2},\\
 &N_R\equiv N_R(\bar{m};g,\delta^2,E_{\textrm{ext}})=\arg\min_N \big({R}_N\big)\big|_{\delta^2}.
 \end{split}
\end{align}
In Section~\ref{EngtoComp}, $E_{\textrm{ext}}=0$ and the two optimal steps coincide, $N_C=N_R$. However, with the additional {energy} cost they start to differ and the step optimisation becomes dependent on the quantity to be minimised. By construction, optimising for the total number of gates will result in a lower complexity at the expense of a higher total {energy} cost,
\begin{equation}\label{tworesources}
R_{N_C} \geq R_{N_R},
\end{equation}
and vice versa, with the equality attained if $E_{\textrm{ext}}=0$. 
\subsubsection{State Preparation}\label{sec:stateprep}
State preparation often aims to cool a thermal state from its initial temperature to a lower one. In our case, the ideal initial state of the system qubit is $\ketbra{0}{0}$. However, the third law of thermodynamics implies that any process cannot reach zero temperature, corresponding to a pure state, with finite resource \cite{Masanes_2017}, and so the cooled state can only lie in the vicinity of $\ketbra{0}{0}$. For the implementation, we adopt the technique of dynamic cooling \cite{Vazirani,Bassman_Oftelie_2024}. Consider $M_s$ identical qubits, with one target system and $M_s-1$ auxiliary qubits. They are governed by the same Hamiltonian, $H_i=-\hbar\omega_0\sigma_z^{(i)}/2$, $i=1,2,...,M_s$. Their initial state is a product of thermal states with environmental temperature $T_0$:
\begin{equation}\label{coolinitial}
    \rho_\beta=\bigotimes _{i=1}^{M_s}\left(\frac{e^{-\beta H_i}}{Z(\beta)}\right),
\end{equation}
where $\beta=1/k_BT_0$, $k_B$ being Boltzmann's constant, and $Z(\beta)$ is the corresponding partition function. A unitary operation $V_{sp}$ then acts on all qubits and the auxiliary ones are discarded afterwards, leaving the system qubit in a state with new temperature $T<T_0$. It is shown \cite{Bassman_Oftelie_2024} that in the low temperature regime, the minimal temperature the cooling can reach is
\begin{equation}\label{cooling}
T\approx \frac{2T_0}{M_s},\quad\text{if }k_BT_0\ll \hbar\omega_0\,.
\end{equation}
{The energy cost of state preparation is again interpreted as in Section~\ref{sec:energynotion}. Here the initial state $\rho_\beta$ is considered free of cost \cite{Gallego_2016,Lostaglio_2019}, as it is in thermal equilibrium with the environment.  The cost $W$ comes from the energy change of the qubits induced by the non-energy-conserving unitary $V_{sp}$,  and the ensuing free operation consists of discarding the auxiliary qubits. Eq.~\eqref{def:energy} thus reads 
\begin{equation}\label{eq:W}
W=\Tr\bigg[\Big(\sum_i H_i\Big)(V_{sp}\rho_\beta V_{sp}^\dag-\rho_\beta)\bigg].    
\end{equation}
According to the model of Figure~\ref{fig:energynotion}, $W$ can be interpreted as the free energy drawn from an external battery \cite{Clivaz_2019} to implement the cooling unitary.} Ref.~\cite{Bassman_Oftelie_2024} shows that {$W$ in Eq.~\eqref{eq:W}} is an extensive quantity and the {energy cost} per qubit in the thermodynamic limit is
\begin{equation}\label{workdonecooling}
\bar{w}=\lim_{M_s\rightarrow\infty}\frac{W}{M_s}=\frac{\omega_0}{2\omega}\frac{\tanh(\frac{1}{2\xi})
    }{e^{\frac{1}{\xi}}+1}, \quad \xi=\frac{k_BT_0}{\hbar\omega_0},
\end{equation}
where the energy unit remains the photonic energy of the EM field. As seen shortly, in practice only a small number of cooling qubits are needed for {energy} optimisation and so the thermodynamic limit does not apply. For small $T_0$ we can nevertheless approximate $\bar{w}$ as the {energy change of} each qubit. The external cost per round for state preparation is then 
\begin{equation}\label{Eextcool}
E_{\textrm{ext}}^{{\textrm{(preparation)}}}=\bar{w}\cdot M_s.
\end{equation}
\begin{figure}[t]
\includegraphics[width=0.45\textwidth]{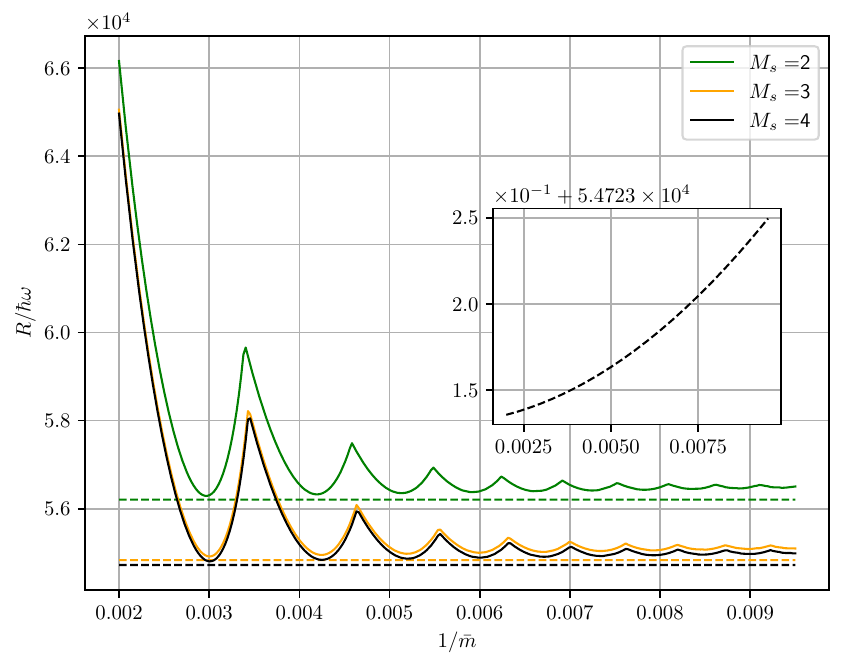}
    \caption{Plots of the total {energy cost (in units of $\hbar\omega$)} of {implementing the phase-encoding gate (see Figure~\ref{plot:QPEEnergy}), now combined with that of state preparation for initialising the probe qubit}, with $g=2.5$, $\xi=0.2$, $\delta^2=10^{-4}$ and the error represented by $1/\bar{m}$.   The number of cooling qubits used per sequence, $M_s$, is also varied to show that small amount of cooling suffices to approach the ideal limit. Solid lines are computed by Eq.~\eqref{totalcostext}, while dashed ones \eqref{QPEresourceT} approximate the cost through the raw complexity. Contrary to the plateau in Figure~\ref{plot:QPEEnergy}, the total {energy} cost ends up increasing with larger implementation error as shown in the inset, since a larger error, or smaller $\bar{m}$, leads to more repetitions (see Eq.~\eqref{optimalstepapprox}) and hence a larger cooling cost. This corresponds to the behaviour indicated by the solid curve in Figure~\ref{figure:resource}(c). However, the increment is negligible {and only observable by the zoom-in in the inset} as the cost of cooling is much smaller than that of gate implementation.}\label{coolingexample}
\end{figure}
This quantity later turns out to be much smaller than the cost of gate implementation for state-of-the-art technology. In this subsection we will thus ignore the distinction between complexity and energy minimisation introduced in Section~\ref{sec:optimalstepmod}. 

After cooling, each sequence now starts with the initial state
\begin{equation}\label{thermalstate}
\rho_0(T)=\frac{e^{-\beta H_i}}{Z(\beta)}=\frac{1+\gamma(T)}{2}\ketbra{0}{0}+\frac{1-\gamma(T)}{2}\ketbra{1}{1},
\end{equation}
where $\gamma(T)=\tanh{\left(\frac{\hbar\omega_0}{2k_BT}\right)}$ characterises the closeness to $\ketbra{0}{0}$ as the Bloch vector of $\rho_0$ is $[0,0,\gamma(T)]^\intercal$. Derivations in Appendix~\ref{appendix:computingQFI} imply that the leading correction to the QFI due to the non-ideal initial state is a multiplicative factor of $[\gamma(T)]^2$. With Eq.~\eqref{cooling} this yields modification on Eqs.~\eqref{QFIapproximated}--\eqref{approxQFIComp} as  
\onecolumngrid
\begin{align}\label{coolscaling}
\begin{split}
F_N(\bar{m};g)&\xrightarrow{\text{state preparation}}F_N(\bar{m};g,M_s)\approx\left[\gamma\left(\frac{2T_0}{M_s}\right)\right]^{2}\cdot F_N(\bar{m};g);\\
N_{\textrm{opt}}(\bar{m};g)&\xrightarrow{\text{state preparation}}N_{\textrm{opt}}(\bar{m};g,M_s)\approx N_{\textrm{opt}}(\bar{m};g);\\
q_{N_{\textrm{opt}}}(\bar{m};g,\delta^2)&\xrightarrow{\text{state preparation}} q_{N_{\textrm{opt}}}(\bar{m};g,\delta^2,M_s)\approx\left[\gamma\left(\frac{2T_0}{M_s}\right)\right]^{-2} q_{N_{\textrm{opt}}}(\bar{m};g,\delta^2);\\
c(\bar{m};g,\delta^2)&\xrightarrow{\text{state preparation}} c(\bar{m};g,\delta^2,M_s)\approx\left[\gamma\left(\frac{2T_0}{M_s}\right)\right]^{-2}c(\bar{m};g,\delta^2).
\end{split}
\end{align}
\twocolumngrid
The total {energy} cost is computed through Eq.~\eqref{totalcostext}, with $N$ determined by the minimisation in \eqref{QPEcomplexity} and $E_{\textrm{ext}}$ given by Eq.~\eqref{Eextcool}. For small $\bar{m}$, this can be approximated through the raw complexity: using Eq.~\eqref{coolscaling} on Eq.~\eqref{totalcostext} leads to,
\begin{equation}\label{QPEresourceT}
    R(\bar{m};g,\delta^2,M_s)\approx \frac{e}{\delta^2}\cdot\frac{\Delta(g)^2}{\bar{m}^2}\cdot\frac{\bar{w}M_s+\frac{\bar{m}^2}{\Delta(g)}}{\left[\tanh\left(\frac{M_s}{4\xi }\right)\right]^2}.
\end{equation}

A characteristic value of $\xi$ for contemporary quantum technologies based on techniques such as superconducting and ion trap qubits is approximately $0.2$ \cite{Superconducting,TrappedIon}; Eq.~\eqref{workdonecooling} then yields $\bar{w}\approx 0.003$, where we take advantage of the resonance condition $\omega\approx\omega_0$. Within this range, Figure~\ref{coolingexample} shows that the total {energy} cost quickly converges to a minimal value when more cooling qubits are consumed per sequence. From Eq.~\eqref{QPEresourceT} we anticipate the total cost to eventually increase with $M_s$ --- complexity reduction brought by further cooling is overpowered by its energy cost. However, this does not manifest until $M_s$ grows significantly large, since the relative magnitude between the cost of state preparation and gate implementation is as small as 
\begin{equation}\label{coolingmagnitude}
\frac{\text{cost of states}}{\text{cost of gates}}\sim \frac{\bar{w}M_s\Delta(g)}{\bar{m}^2}\sim 10^{-7}M_s.
\end{equation}

\subsubsection{Measurement}\label{sec:mmt}
To implement the measurement procedure we adopt the pointer model \cite{von2018mathematical,Alexandre_Brasil_2015}: the system qubit is coupled to a pointer qubit through a CNOT gate controlled by the system, the pointer is measured with respect to the {optimal POVM that maximises the CFI, reported in Eq.~\eqref{eqn:optimalPOVM},} and the outcome should follow the statistics of the system state. The pointer thus acts as a measuring device. {Let it be governed by the Hamiltonian $-\hbar\omega_1\sigma_z/2$, $\omega_1$ being the transition frequency.}

For perfect measurement, we want to initialise the pointer in $\ketbra{0}{0}$. However, again, a pure state requires an infinite amount of resource to prepare, and so does an ideal projective measurement \cite{Guryanova2020}. The resulting non-ideal measurement affects the QPE performance \cite{Len_2022,Kurdzia_ek_2023}. In Appendix~\ref{appendix:mmt} we evaluate the CFI from a mixed pointer state prepared through the dynamic cooling protocol. Despite the imperfection, as observed in Section~\ref{sec:stateprep}, in current setups the energy cost of cooling is negligible compared to that of the phase-encoding gate implementation for energy optimisation. We will therefore assume the initial pointer state is $\ketbra{0}{0}$ for free. For an $N$-step sequence the system-pointer state prior to the correlating gate is then of the form $\rho_N^{(S)}\otimes\ketbra{0}{0}^{(P)}$, where superscripts are used to label the two parts. {This is taken as the free initial state in the bookkeeping model of Figure~\ref{fig:energynotion} when formalising the energy cost, while measurement on the pointer{~\cite{MMTnote}} and discarding of the system qubit amount to the free operations at the end. The cost in this case comes from the correlating CNOT gate: the resulting energy increase it causes on the pointer qubit (since the system qubit acts as the control, its energy is not changed) is given by}
\begin{align}
E_{\textrm{ext}}^{{\textrm{(measurement)}}}=&\Tr\bigg[\left(\mathcal{I}^{(S)}\otimes \frac{-\omega_1}{2\omega}\sigma_z^{(P)}\right)\cdot \\
&\!\!\!\!\!\!\!\!\!\!\!\!\!\!\!\!\!\!\!\!\!\!\!\!\!\left(\textrm{CNOT}\big(\rho_N^{(S)}\otimes\ketbra{0}{0}^{(P)}\big)\textrm{CNOT}^\dag -\rho_N^{(S)}\otimes\ketbra{0}{0}^{(P)}\right)\bigg]\nonumber\\=&\bra{1}\rho_N^{(S)}\ket{1}\cdot\frac{\omega_1}{\omega}\,,   \nonumber
\end{align}
in units of the photonic energy. 
Here $E_{\textrm{ext}}^{{\textrm{(measurement)}}}$ depends on the system state just before the CNOT gate is applied. For both simplicity and generality, we will instead use the loose yet constant upper bound on the correlating cost, namely the gap between the two energy levels of the pointer qubit: 
\begin{equation}\label{mmtperround}
E_{\textrm{ext}}^{{\textrm{(measurement)}}} \cong\frac{\omega_1}{\omega}.
\end{equation}

Both types of {total energy} cost in Eq.~\eqref{tworesources} are plotted in Figure~\ref{figure:mmtcost} for a range of $\frac{\omega_1}{\omega}$ chosen for illustrative purpose. {Note that they are within the approximation range according to Eq.~\eqref{coolingmagnitude} --- replacing $\omega_0$ with $\omega_1$ when computing $\bar w$ in Eq.~\eqref{workdonecooling} --- to validate the omission of the state preparation (cooling) cost for the pointer.} 

\begin{figure}[t]
\includegraphics[width=0.45\textwidth]{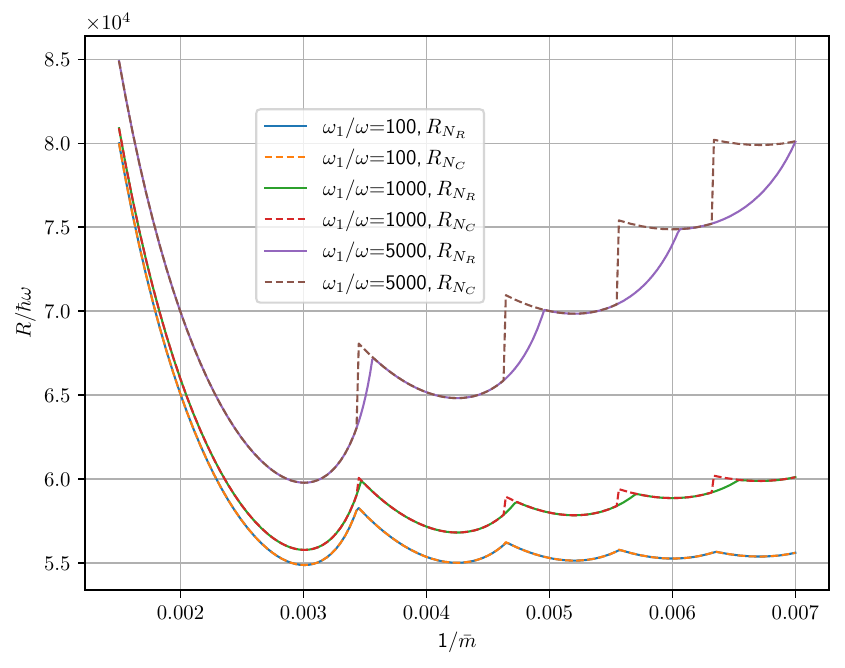}
    \caption{{Plots of the combined total energy cost of measurement and gate implementation (in units of $\hbar\omega$).} The fixed parameters are $g=2.5$, $\delta^2=10^{-4}$. Solid and dashed lines correspond to when the total {energy} cost and the number of gates are minimised, respectively amounting to  $R_{N_R}$ and $R_{N_C}$; see Eqs.~\eqref{totalcostext}--\eqref{tworesources}, with the measurement cost $E_{\textrm{ext}}$ given by Eq.~\eqref{mmtperround}, or $\omega_1/\omega$. The two costs coincide when they involve the same number of complete sequences, $Q_N$, since then their measurement costs, $E_{\textrm{ext}}\cdot (Q_N+1)$, are the same and so minimising the total energy cost and only the cost of gate implementation are equivalent. Otherwise, a gap opens up between $R_{N_C}$ and $R_{N_R}$ when the numbers of full sequences for each differ. This happens when the measurement cost of an extra sequence outweighs the cost of the gates it saves for complexity minimisation. Observe that, the larger and the smaller $E_{\textrm{ext}}$ and $\bar{m}$ are, respectively, the larger the gap grows between the two energy costs and the choice regarding which quantity to optimise becomes more important.}\label{figure:mmtcost}
\end{figure}

The gap between the two costs emphasises the importance of deciding the optimisation priority between complexity and energy, a task closely related to practical constraints. Finally, recall Eq.~\eqref{optimalpoint} and its preceding arguments. Due to the additional cost each round, the sweet spot now becomes a turning point of the total {energy} cost as a function of the implementation error. This characteristic  --- reminiscent of the qualitative behaviour displayed by the solid curve in Figure~\ref{figure:resource}(c) --- further justifies the spot as a feasible candidate for complexity-energy co-optimisation. Interestingly, such an optimal working point is uniquely identified regardless of which minimisation is chosen for the number of steps, as the gap between $R_{N_C}$ and $R_{N_R}$ only begins to open up at larger error values.

\subsection{{Discussion}}\label{sec:discussion}

\begin{figure*}[th]
    \centering
    \begin{tikzpicture}[scale=0.5]
        \draw (-12.7,-9.5) node{$T_0$};
        \filldraw (-13.5,-10) circle (2pt);
        \draw (-13.5,-10)--(-12,-10);
        \filldraw (-13.5,-10.5) circle (2pt);
        \draw (-13.5,-10.5)--(-12,-10.5);
        \filldraw (-13.5,-11.5) circle (2pt);
        \draw (-13.5,-11.5)--(-12,-11.5);
        
        \filldraw[black] (-12.7,-10.7) circle (1pt);
        \filldraw[black] (-12.7,-11) circle (1pt);
        \filldraw[black] (-12.7,-11.3) circle (1pt);

        \draw (-12,-12) rectangle (-10,-9.5)node[anchor= center, midway]{\Large $V_{sp}$};
        \draw (-10,-11) -- (-9,-11);
        \draw (-9.2,-11.3) -- (-8.8,-10.7);
        \draw (-10,-10) -- (-8,-10);
        \draw (-8,-11) rectangle (-5,-8) node[anchor= center, midway]{\large $e^{-iH_St}$};
        \draw (-9,-8.5) -- (-8,-8.5);
        \draw (-9.5,-8.5)node[]{\large $|\bar{m}\rangle$};
        \draw (-5,-10) -- (-4,-10);
        \filldraw[black] (-3.5,-10) circle (1pt);
        \filldraw[black] (-3.0,-10) circle (1pt);
        \filldraw[black] (-2.5,-10) circle (1pt);
        \draw (-2.3,-10) -- (-1.3,-10);
        \draw (-1.3,-11) rectangle (1.7,-8)node[anchor= center, midway]{\large $e^{-iH_St}$};
        \draw (-2.3,-8.5) -- (-1.3,-8.5);
        \draw (-2.8,-8.5)node[]{\large $|\bar{m}\rangle$};

        \draw (1.7,-10)--(6.2,-10);
        \draw (6,-10.3)--(6.4,-9.7);
        \filldraw (5.2,-10) circle (5pt);

        \filldraw (0.7,-11.5) circle (2pt);
        \draw (0.7,-11.5)--(2.2,-11.5);
        \filldraw (0.7,-12.7) circle (2pt);
        \draw (0.7,-12.7)--(2.2,-12.7);
        \draw (2.2,-13.2) rectangle (4.2,-10.7)node[anchor=center,midway]{\Large $V_{m} $};
        \draw (4.2,-12.2)--(5.2,-12.2);
        \draw (5,-12.5) -- (5.4,-11.9);
        \draw (4.2,-11.5) -- (6 ,-11.5);
        
        \draw (5.2,-11.5) circle (5pt);
        \draw (5.2,-11.68) -- (5.2,-9.82);
        \node[meter] (meter) at (6.72,-11.5) {};
        
        \draw[decorate,decoration={brace,amplitude=5pt,mirror}] (0.3,-11.5) -- (0.3,-12.7) node[midway,xshift=-15pt, rotate=90]{ $M_m$ };

        \filldraw[black] (1.45,-11.8) circle (1pt);
        \filldraw[black] (1.45,-12.1) circle (1pt);
        \filldraw[black] (1.45,-12.4) circle (1pt);
         
        \draw[decorate,decoration={brace,amplitude=10pt,mirror}] (-6.5,-21.5) -- (-0.5,-21.5) node[midway,yshift=-25pt]{\Large $N$ steps};
        \draw[decorate,decoration={brace,amplitude=5pt,mirror}] (-14,-10) -- (-14,-11.5) node[midway,xshift=-15pt, rotate=90]{ $M_s$ qubits};

        \filldraw[black] (-11,-13.5) circle (1.5pt);
        
        \filldraw[black] (-11,-14) circle (1.5pt);
        \filldraw[black] (-11,-14.5) circle (1.5pt);
        \filldraw[black] (-11,-15) circle (1.5pt);
        \filldraw[black] (-11,-15.5) circle (1.5pt);
        \filldraw[black] (-11,-16) circle (1.5pt);
        \filldraw[black] (-11,-16.5) circle (1.5pt);

           \filldraw[black] (-6.5,-13.5) circle (1.5pt);
        \filldraw[black] (-6.5,-14) circle (1.5pt);
        \filldraw[black] (-6.5,-14.5) circle (1.5pt);
        \filldraw[black] (-6.5,-15) circle (1.5pt);
        \filldraw[black] (-6.5,-15.5) circle (1.5pt);
        \filldraw[black] (-6.5,-16) circle (1.5pt);
        \filldraw[black] (-6.5,-16.5) circle (1.5pt);

             \filldraw[black] (-3,-13.5) circle (1.5pt);
        \filldraw[black] (-3,-14) circle (1.5pt);
        \filldraw[black] (-3,-14.5) circle (1.5pt);
        \filldraw[black] (-3,-15) circle (1.5pt);
        \filldraw[black] (-3,-15.5) circle (1.5pt);
        \filldraw[black] (-3,-16) circle (1.5pt);
        \filldraw[black] (-3,-16.5) circle (1.5pt);

             \filldraw[black] (5.2,-13.5) circle (1.5pt);
        \filldraw[black] (5.2,-14) circle (1.5pt);
        \filldraw[black] (5.2,-14.5) circle (1.5pt);
        \filldraw[black] (5.2,-15) circle (1.5pt);
        \filldraw[black] (5.2,-15.5) circle (1.5pt);
        \filldraw[black] (5.2,-16) circle (1.5pt);
        \filldraw[black] (5.2,-16.5) circle (1.5pt);

        \filldraw (-13.5,-20) circle (2pt);
        \draw (-13.5,-20)--(-12,-20);
        \filldraw (-13.5,-20.5) circle (2pt);
        \draw (-13.5,-20.5)--(-12,-20.5);
        \filldraw (-13.5,-21.5) circle (2pt);
        \draw (-13.5,-21.5)--(-12,-21.5);
        
        \filldraw[black] (-12.7,-20.7) circle (1pt);
        \filldraw[black] (-12.7,-21) circle (1pt);
        \filldraw[black] (-12.7,-21.3) circle (1pt);

        \draw (-12,-22) rectangle (-10,-19.5)node[anchor= center, midway]{\Large $V_{sp}$};
        \draw (-10,-21) -- (-9,-21);
        \draw (-9.2,-21.3) -- (-8.8,-20.7);
        \draw (-10,-20) -- (-8,-20);
        \draw (-8,-21) rectangle (-5,-18) node[anchor= center, midway]{\large $e^{-iH_St}$};
        \draw (-9,-18.5) -- (-8,-18.5);
        \draw (-9.5,-18.5)node[]{\large $|\bar{m}\rangle$};
        \draw (-5,-20) -- (-4,-20);
        \filldraw[black] (-3.5,-20) circle (1pt);
        \filldraw[black] (-3.0,-20) circle (1pt);
        \filldraw[black] (-2.5,-20) circle (1pt);
        \draw (-2.3,-20) -- (-1.3,-20);
        \draw (-1.3,-21) rectangle (1.7,-18)node[anchor= center, midway]{\large $e^{-iH_St}$};
        \draw (-2.3,-18.5) -- (-1.3,-18.5);
        \draw (-2.8,-18.5)node[]{\large $|\bar{m}\rangle$};

         \draw (1.7,-20)-- (6.2,-20);
         \draw(6,-20.3)--(6.4,-19.7);
        \filldraw (5.2,-20) circle (5pt);

        \filldraw (0.7,-21.5) circle (2pt);
        \draw (0.7,-21.5)--(2.2,-21.5);
        \filldraw (0.7,-22.7) circle (2pt);
        \draw (0.7,-22.7)--(2.2,-22.7);
        \draw (2.2,-23.2) rectangle (4.2,-20.7)node[anchor=center,midway]{\Large $V_{m}$};
        \draw (4.2,-22.2) -- (5.2,-22.2);
        \draw (5,-22.5) -- (5.4,-21.9);
        \draw (4.2,-21.5) -- (6 ,-21.5);
        \draw (5.2,-21.5) circle (5pt);
        \draw (5.2,-21.68) -- (5.2,-19.82);
        \node[meter] (meter) at (6.72,-21.5) {};

        \filldraw[black] (1.45,-21.8) circle (1pt);
        \filldraw[black] (1.45,-22.1) circle (1pt);
        \filldraw[black] (1.45,-22.4) circle (1pt);

\begin{scope}[yshift=-7cm]
        \filldraw[black] (-12.7,-20.7) circle (1pt);
        \filldraw[black] (-12.7,-21) circle (1pt);
        \filldraw[black] (-12.7,-21.3) circle (1pt);

        \draw (-12,-22) rectangle (-10,-19.5)node[anchor= center, midway]{\Large $V_{sp}$};
        \draw (-10,-21) -- (-9,-21);
        \draw (-9.2,-21.3) -- (-8.8,-20.7);
        \draw (-10,-20) -- (-8,-20);
        \draw (-8,-21) rectangle (-5,-18) node[anchor= center, midway]{\large $e^{-iH_St}$};
        \draw (-9,-18.5) -- (-8,-18.5);
        \draw (-9.5,-18.5)node[]{\large $|\bar{m}\rangle$};
        \draw (-5,-20) -- (-4,-20);
        \filldraw[black] (-3.5,-20) circle (1pt);
        \filldraw[black] (-3.0,-20) circle (1pt);
        \filldraw[black] (-2.5,-20) circle (1pt);
        \draw (-2.3,-20) -- (-1.3,-20);
        \draw (-1.3,-21) rectangle (1.7,-18)node[anchor= center, midway]{\large $e^{-iH_St}$};
        \draw (-2.3,-18.5) -- (-1.3,-18.5);
        \draw (-2.8,-18.5)node[]{\large $|\bar{m}\rangle$};

         \draw (1.7,-20)-- (6.2,-20);
         \draw(6,-20.3)--(6.4,-19.7);
        \filldraw (5.2,-20) circle (5pt);

        \filldraw (0.7,-21.5) circle (2pt);
        \draw (0.7,-21.5)--(2.2,-21.5);
        \filldraw (0.7,-22.7) circle (2pt);
        \draw (0.7,-22.7)--(2.2,-22.7);
        \draw (2.2,-23.2) rectangle (4.2,-20.7)node[anchor=center,midway]{\Large $V_{m}$};
        \draw (4.2,-22.2) -- (5.2,-22.2);
        \draw (5,-22.5) -- (5.4,-21.9);
        \draw (4.2,-21.5) -- (6 ,-21.5);
        \draw (5.2,-21.5) circle (5pt);
        \draw (5.2,-21.68) -- (5.2,-19.82);
        \node[meter] (meter) at (6.72,-21.5) {};

        \draw[decorate,decoration={brace,amplitude=10pt,mirror}] (-6.5,-21.5) -- (-0.5,-21.5) node[midway,yshift=-25pt]{\Large $N_0$ steps};

        \filldraw[black] (1.45,-21.8) circle (1pt);
        \filldraw[black] (1.45,-22.1) circle (1pt);
        \filldraw[black] (1.45,-22.4) circle (1pt);

        \filldraw (-13.5,-20) circle (2pt);
        \draw (-13.5,-20)--(-12,-20);
        \filldraw (-13.5,-20.5) circle (2pt);
        \draw (-13.5,-20.5)--(-12,-20.5);
        \filldraw (-13.5,-21.5) circle (2pt);
        \draw (-13.5,-21.5)--(-12,-21.5);
        
        \filldraw[black] (-12.7,-20.7) circle (1pt);
        \filldraw[black] (-12.7,-21) circle (1pt);
        \filldraw[black] (-12.7,-21.3) circle (1pt);
\end{scope}
        
        \draw[decorate,decoration={brace,amplitude=5pt}] (8,-10) -- (8,-20) node[midway,xshift=20pt, rotate=270]{\Large $Q_{N}$ rounds};

    \end{tikzpicture}
    \caption{Full circuit diagram of the QPE protocol for which a trade-off between complexity and energy is established in this work.}
    \label{fig:circuitcool}
\end{figure*}

Figure~\ref{fig:circuitcool} summarises the QPE protocol constructed {in this Section}. For a fixed number of photons $\bar m$ spent on each gate, we have determined the minimal number of gates needed, or the complexity, and their spatial arrangement to reach a desired lower bound on the estimator variance. We have also analysed the corresponding total energy cost and identified a sweet spot for complexity-energy co-optimisation while varying $\bar m$.

{As mentioned at the end of Section~\ref{section:introduction}, our energy analysis on the QPE protocol fits naturally into the MNR framework \cite{AlexiaQEI,AlexiaStack}. In Eq.~\eqref{MNREff}, we can identify the {\em metric} as the total QFI of the sequential protocol, constrained by the target bound $1/\delta^2$, while the {\em resource} is the total energy cost $R$ computed so far. The {\em noise} in our case comes from the quantum fluctuation of the EM field. Therefore, for each fixed $\delta^2$, maximising the efficiency~\eqref{MNREff}, $\eta=\frac{1}{\delta^2R}$, is equivalent to minimising the total energy cost. 

When only the cost of the sequential gate implementation is factored in as in Section~\ref{EngtoComp}, the minimal cost corresponds to the saturation point~\eqref{optimalpoint}. In this case the efficiency becomes independent of the metric: $\eta=\frac{1}{\delta^2R_0}=\frac{1}{e\Delta(g)}$. Once the other sources of cost --- including but not limited to the ones analysed in Section~\ref{section:othercost} --- kick in, the total energy cost increases and the metric dependence of the efficiency starts to show. Therefore, within the MNR framework the QPE protocol constructed in this Section has an efficiency upper bounded by
\begin{equation}\label{MNRefficiencyy}
    \eta \leq \frac{1}{e\Delta(g)}.
\end{equation}
From Eq.~\eqref{QFIapproximated}, $\Delta(g)$ is a monotonically increasing function of $g$ and so, the larger the unknown parameter $g$, the less energy-efficient the QPE protocol is.} 
{The primary objective of our analysis is the joint optimisation of
$C(\epsilon)$ and $E(\epsilon)$ so as to minimise
$R(\epsilon)$ at fixed metrological performance.
Eq.~\eqref{MNRefficiencyy} recasts the resulting optimum within
the MNR framework and quantifies the intrinsic efficiency of the
quantum implementation considered here. More broadly, the MNR framework can incorporate additional platform-dependent contributions, such as the operation of control equipment and the effects of environmental noise, through the corresponding extensions of the total cost and performance metric. The bound in Eq.~\eqref{MNRefficiencyy} therefore provides a fundamental benchmark against which the efficiency of more complete experimental implementations can be assessed.}

{The energy plateau observed in Fig.~\ref{plot:QPEEnergy} motivates us to ask if the asymptotic scaling of the total energy cost with sufficiently large implementation error can be inferred from just a small set of parameters that are well-defined for all physical realisations of the QPE protocol, so that qualitative conclusions can be drawn in a universal fashion. This can serve for both theoretical and engineering purposes in designing energy-optimal physical realisations of sequential QPE tasks. Here we undertake a preliminary investigation and determine a condition for the energy plateau to exist. 

First, we state our general assumptions, which hold in particular for the optical implementation  analysed in this Section:
\begin{enumerate}
    \item We will stay in the semi-classical regime so that the implementation error $\epsilon$ can be treated as a small parameter;
    \item Meanwhile, $\epsilon$ shall also be large enough such that the raw complexity well approximates the true complexity;
    \item Recalling the qualitative form of the QFI in Eq.~\eqref{QFIGeneral},
    \begin{equation*}
        F_N=N^2e^{-f(N,\epsilon)},
    \end{equation*}  
    the exponent $f$ will be assumed to grow no more than polynomially as a function of the number of steps $N$;
    \item\label{itm:energy} The implementation energy of each phase-encoding gate will be assumed to follow an inverse power law, so that $E(\epsilon)\sim\epsilon^{-\tau}$ for some $\tau>0$. 
\end{enumerate}

In the ideal limit of vanishing implementation error, $\epsilon=0$, we reach the Heisenberg limit of a sequential QPE strategy~\cite{Maccone2013}, $F_N\propto N^2$. Without loss of generality, set $f(N,0)=0$ so that the constant of proportionality is $1$. Then, $f$ admits a power series expansion with respect to $\epsilon \rightarrow 0$ with some leading order $\nu>0$:
\begin{equation*}
f(N,\epsilon) \approx g(N)\epsilon^{\nu},
\end{equation*}
for some polynomially-bounded function $g(N)$. The optimal step $N_{\textrm{opt}}$ for the raw complexity~\eqref{rawcomplexity} is determined by minimising the quantity $Nf(N,\epsilon)$, leading to 
\begin{equation*}
N_{\textrm{opt}}\frac{\partial f(N_{\textrm{opt}},\epsilon)}{\partial N}=1\quad\longrightarrow\quad N_{\textrm{opt}}\,g'(N_{\textrm{opt}}) \approx \epsilon^{-\nu}.
\end{equation*}
Let $g(N)$ admit a power series expansion with leading order $q>0$, so that $g(N) \approx c_qN^q$ as $N\rightarrow\infty$. The optimal step and the corresponding raw complexity can then be expressed as
\begin{align*}
N_{\textrm{opt}}&\approx(c_qq\epsilon^{\nu})^{-1/q},\\
c(\epsilon;\delta^2)&=\frac{1}{\delta^2N_{\textrm{opt}}e^{-f(N_{\textrm{opt}},\epsilon)}}\approx \delta^{-2}(ec_qq\epsilon^\nu)^{1/q}.
\end{align*}
Within the regime of validity of our assumptions, the total energy cost~\eqref{noisyRes} then behaves as
\begin{equation}\label{eq:totalenergygeneral}
R(\epsilon;\delta^2)=E(\epsilon)\times c(\epsilon;\delta^2)\sim\mathcal{O}(\epsilon^{\frac{\nu}{q}-\tau}).
\end{equation}

Observe that $\nu=q\tau$ and $\nu>q\tau$ correspond to the dashed and solid curve in Figure~\ref{figure:resource}, respectively, where the total energy cost either eventually plateaus or exhibits a global minimum; on the other hand,  $\nu<q\tau$ indicates that the total energy cost vanishes asymptotically, which is incompatible with the stated assumptions. 

The example in Section~\ref{EngtoComp} has $\epsilon=1/\bar m$, and so, according to Eq.~\eqref{QFIapproximated}, $f(N,\epsilon)=-2N\log(1-\Delta(g)\epsilon/2)$ with $\nu=1$, $q=1$ and $c_q=\Delta(g)$. The energy cost per gate is proportional to the photon number, $\tau=1$. Thus $q\tau=1=\nu$ yields the energy plateau as expected. Including the additional costs associated with state preparation and measurement effectively reduces $\tau$: this breaks the plateau and pushes the total energy (\ref{eq:totalenergygeneral}) in the regime $\nu>q\tau$, leading to the appearance of the turning point as identified in Section~\ref{section:othercost}.}

{Finally, note that Assumption~\ref{itm:energy} captures the physically relevant situation in which increasingly accurate gates require increasingly energetic control fields. Consequently, the implementation cost diverges as $\epsilon\rightarrow0$, ruling out the ideal limit as the energy-optimal operating point and allowing a sweet spot to occur at finite error, as illustrated in Fig.~\ref{figure:resource}. This behaviour is not specific to QPE: inverse-power-law relations between control energy and implementation error arise broadly when qubits are driven by EM fields~\cite{Banacloche2002,Ikonen_2017}, including in optical~\cite{Ralph_2006,Flamini_2018} and trapped-ion~\cite{HAFFNER_2008,Chen_2021} platforms.

This interpretation is also consistent with the energy accounting framework introduced in Section~\ref{sec:energynotion}. The external source prepares the control field, which then supplies the energy required to implement the desired qubit transformation. When the target unitary does not conserve the qubit energy, approaching an ideal implementation generally requires an increasing energetic investment in the control operation~\cite{Chiribella2021,giulionote}. Combining this error-energy relation with the corresponding error dependence of the protocol complexity, as in Eq.~\eqref{eqn:ResourceGeneral}, leads to the same form of complexity-energy co-optimisation for a broader class of quantum information-processing tasks.}

\section{Concluding Remarks and Outlook}\label{section:outlook}

In this work, we have established a framework for analysing and quantifying the trade-off between gate complexity and energy cost in quantum processes composed of elementary building blocks. We have applied our framework to the case study of a sequential quantum phase estimation protocol and identified physical conditions for the joint optimisation of complexity and energy, given a target metric benchmarking the desired estimation precision. Our analysis rests on a consistent characterisation of energy cost across the different components of the protocol, encompassing input probe-state preparation, phase-encoding gate implementation, and output measurement.

{Gravitational-wave detection provides a prominent example of quantum-enhanced optical metrology in which energetic and temporal resources must be carefully balanced~\cite{LIGO}. Laser interferometers detect the minute relative displacements of suspended mirrors induced by passing gravitational waves, using intense optical fields whose sensitivity is ultimately constrained by quantum noise, including photon shot noise~\cite{Rafal2013,Pitkin_2011,Demkowicz_Dobrza_ski_2015}. In such settings, increasing the optical energy can improve precision but also raises the energetic cost of operating the detector. Our framework provides a natural starting point for analysing this type of trade-off between estimation performance, interrogation time, and optical energy expenditure.}

As mentioned in the Introduction, all components other than the phase-encoding channel~\eqref{QFIChannel} constitute an estimation strategy for the metrology task. A strategy that obeys causal order can be represented by a quantum comb~\cite{taranto2025higherorderquantumoperations,Liu_2024,Qiushi2023,Kurdzialek_2025,ChiribellaComb,andre2026}. Quantum combs encompass structures more general than the protocol considered here, including temporal and spatial correlations, entanglement among probe qubits, memory effects across successive gates, and correlated decoherence noise~\cite{kurdzia2025}. {A particularly important alternative is the parallel strategy, in which several probes interact with the parameter-encoding process simultaneously, employing entanglement among the probes to improve robustness and estimation performance~\cite{shettell2026,Yousefjani_2017a}. These advantages must be balanced against the additional hardware required to prepare, control, and preserve multipartite entanglement~\cite{Monz_2011,Kam_2024}.} Furthermore, causally indefinite strategies, such as the quantum switch~\cite{Chiribella_2013}, have the potential to surpass conventional precision bounds~\cite{Zhao2020,chen2025nonlinearenhancementmeasurementprecision,yin2023experimentalsuperheisenbergquantummetrology}. The energetic cost of quantum-comb strategies for metrology has received comparatively less attention, with Ref.~\cite{chen2025optimalquantummetrologyenergy} providing one recent development; see also Ref.~\cite{Zambon2025} for a complementary study of work extraction from quantum combs. Combining these results with the framework developed here may help determine the complexity-energy trade-off for both components of a quantum metrology task --- the estimation strategy and the parameter-encoding process --- and thereby enable a more complete co-optimisation.

It is also worth remarking that quantum resources can take forms other than energy. For the QPE protocol, {entanglement among the probes can provide the quantum enhancement underlying parallel estimation strategies~\cite{Maccone2013}}, while Ref.~\cite{ahnefeld2025coherenceresourcephaseestimation} investigates how the quality of phase estimation can be characterised in terms of quantum coherence. A more general framework should therefore be capable of accounting consistently for different types of resource cost. This may be achieved by constructing conversion schemes between resources~\cite{Streltsov2015}, which could, for example, relate energy expenditure to the generation of entanglement~\cite{Hackl_2019,B_ny_2018,horodecki2025quantificationenergyconsumptionentanglement} or coherence~\cite{Misra_2016,Marvian_2022}, or by introducing task-dependent hybrid cost functions.

Finally, the building-block principle developed in this work may be applied to other quantum tasks with structures analogous to Eq.~\eqref{sequential}. Whenever a protocol consists of repeated elementary units, reducing the energy invested in each unit will generally increase its implementation error and, consequently, the number of units required to achieve a prescribed performance. In Section~\ref{sec:discussion}, we obtained a qualitative understanding of this trade-off for QPE without relying on the details of the specific physical implementation. The essential ingredient for extending the analysis is a metric $\mathcal{M}(N,\epsilon)$ that depends on the number of elementary units $N$ and their implementation error $\epsilon$, and that quantitatively assesses the performance of the protocol. For the examples discussed in Section~\ref{sec:realintro}, this metric may be the QFI in quantum metrology as adopted in this work, the overlap between the final and target states in Grover's algorithm~\cite{grover1996fastquantummechanicalalgorithm}, or a suitable norm distance between the implemented and desired evolutions in Hamiltonian simulation~\cite{Lloydsim}. The reasoning developed here can therefore be adapted to a broad range of quantum-information-processing tasks.

A comprehensive analysis on practical complexity-energy trade-off relations for existing and upcoming quantum technology primitives will be pivotal for their sustainable development and widespread impact.

\acknowledgments{This work was supported by the Engineering and Physical Sciences Research Council (EPSRC Grants No.~EP/W524402/1, EP/T022140/1, and EP/X010929/1). We acknowledge fruitful discussions with Longyun Chen, Yuxiang Yang, Florian Meier, Victor Montenegro, Hiroyasu Tajima, Francesco Albarelli, Salvatore Tirone and Tommaso Tufarelli.}

\bibliography{ref}
\onecolumngrid
\appendix
\section{{Raw versus True Complexity}}\label{app:rawvstrue}
{In Section~\ref{basic} we introduce the quantum Fisher information (QFI) as a metric quantifying the estimation precision of the QPE protocol. In general, in noisy quantum metrology \cite{Demkowicz_Dobrza_ski_2012,escher2011general} the QFI can be empirically expressed as 
\begin{equation}\label{QFIGeneral}
F_N=N^2e^{-f(N,\epsilon)}
\end{equation}
for some function $f>0$ that grows with both $N$ and $\epsilon$, characterising the information-destroying effect of the error. In the ideal limit, $\epsilon=0$, we have $f(N,0)\sim1$ and we attain the Heisenberg limit for the sequential QPE protocol, $F_N\propto N^2$~\cite{Maccone2013,RafalSeq}. However, in the presence of noise, this scaling often cannot be realised in practice. In particular, for $f\sim \log(N)$ we reach the standard quantum limit, $F_N\sim N$, while, for even stronger implementation error, the QFI eventually decays as $N$ grows, and adding more steps becomes detrimental for the estimation precision. The examples we study in this work correspond to the latter case; see Figure~\ref{plot:QPEAB}(a) and~\ref{plottQFI}.
\begin{figure}[h!]
    \centering
    \includegraphics[width=0.4\linewidth]{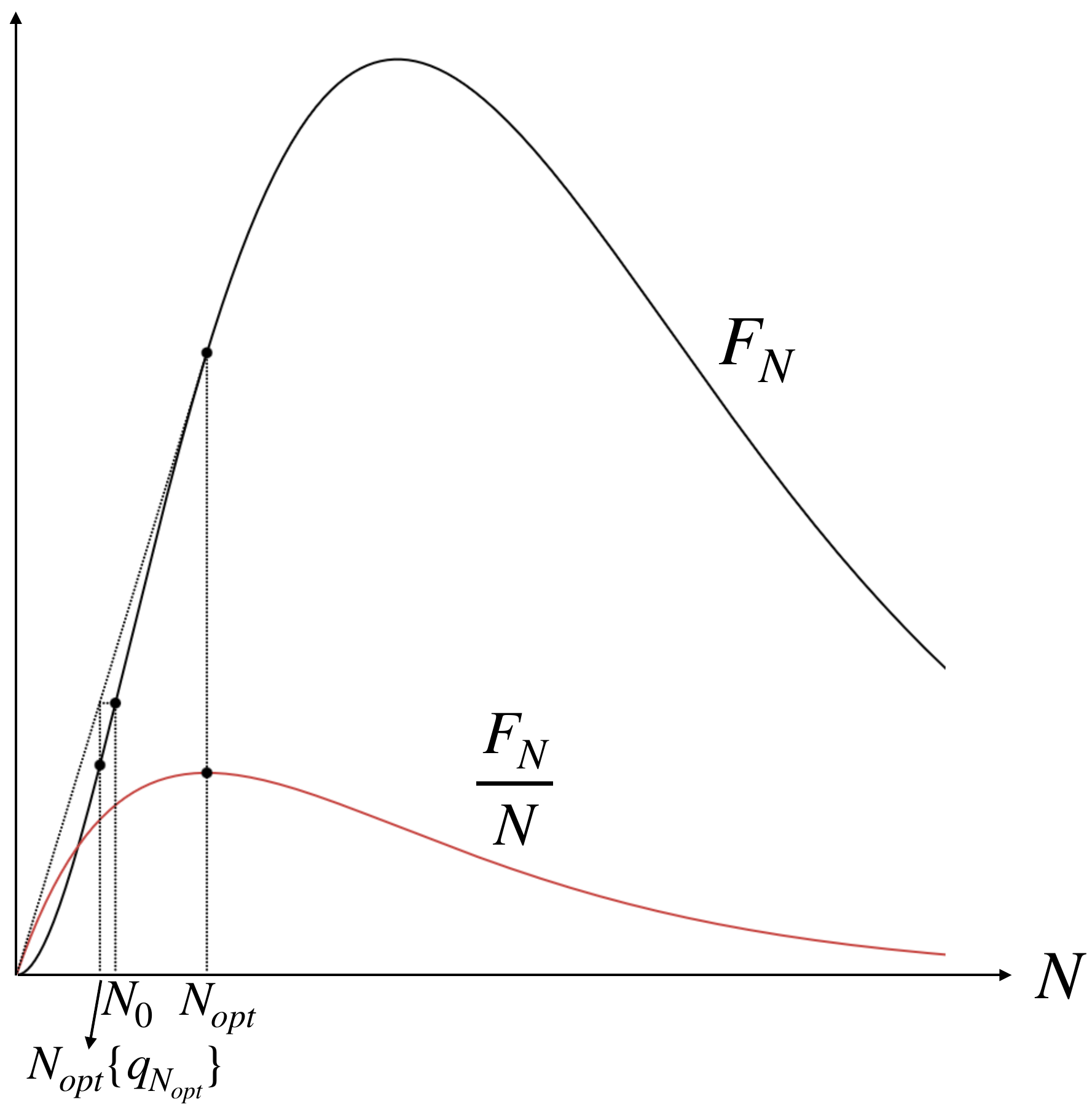}
    \caption{{Typical graph of the QFI $F_N$ and its ratio with $N$ for an asymptotically decaying QFI with respect to the number of steps $N$ in one sequence. Initially, $F_N$ increases quadratically while $F_N/N$ increases linearly. Note that they attain their maximum at different values of $N$;  in particular, $N_{\textrm{opt}}$ maximises $F_N/N$ and leads to the raw complexity in Eq.~\eqref{rawcomplexity}. In the computation there, $N_{\textrm{opt}}\{q_{N_{\textrm{opt}}}\}<N_{\textrm{opt}}$ is taken to be the number of steps in the last sequence, while $N_0$ given by Eq.~\eqref{truelastroundstep} is instead the true number of steps needed to reach the desired QFI. The discrepancy between the two is caused by the nonlinearity of $F_N$ and
    is expected to decrease as the implementation error grows.}  }
    \label{fig:rawtrue}
\end{figure}

With the general qualitative form~\eqref{QFIGeneral}, the difference between the raw and the true complexity becomes clearer to see. For $N$-step sequences, the raw complexity~\eqref{rawcomplexity} is computed assuming the last partial sequence consists of $N\cdot\{q_N\}$ steps, where $\{q_N\}$ is the fractional part of $q_N$, while for the true complexity~\eqref{QPEcomplexity}, the last sequence contains $N_0$ steps that satisfy the condition~\eqref{truelastroundstep}. In general, $F_{N\{q_N\}}\neq F_{N_0}$ unless $F_N$ is linear over $N$, or when $q_N$ is an integer (and so $N_0=\{q_N\}=0$). More precisely, we have
\begin{equation*}
    \frac{N\{q_N\}}{N}=\frac{F_{N_0}}{F_N}=\frac{N_0^2e^{-f(N_0,\epsilon)}}{N^2e^{-f(N,\epsilon)}}>\frac{N_0^2}{N^2},
\end{equation*}
and a rough upper bound on the difference between the actual length of the final sequence needed to reach the desired total QFI and the one as computed by the raw complexity is $N_0-N\{q_N\}<N_0-N_0^2/N\leq N/4$. These quantities are demonstrated in Figure~\ref{fig:rawtrue} for a typical $F_N$ with an eventual decay. Intuitively, as $\epsilon$ increases, the decaying of both $F_N$ and $F_N/N$ becomes faster and the turning point of the latter, $N_{\textrm{opt}}$, decreases. As the difference between the raw and the true complexity is upper bounded by roughly $N_{\textrm{opt}}/4$, and both supposedly increase with increasing error, we anticipate the true complexity to be well approximated by the raw complexity for large enough implementation error $\epsilon$. This is confirmed by the observation made in Figure~\ref{plot:QPEAB}(b) and~\ref{plot:QPEEnergy}.
}

\section{Computation of the Quantum Fisher Information $F_N$}\label{appendix:computingQFI}
To evaluate the QFI resulting from the implementation scheme in Section~\ref{EngtoComp}, we first compute the action of the imperfect quantum channel. Recall Eqs.~\eqref{QFIChannel} and \eqref{QFIChannelBloch}. Write a general qubit state  in its Bloch representation, $\rho=\frac{1}{2}\left(\textbf{I}_2+\textbf{s}\cdot\boldsymbol{\sigma}\right)$, $\textbf{s}$ being the Bloch vector. From Eq.~\eqref{unitarypulse}, the action of the implemented quantum channel on the state can be expressed as
\begin{equation*}
\mathcal{G}_{\bar{m};g}(\rho)=\frac{\textbf{I}_2}{2}+\frac{1}{2}\underbrace{\left(\int_0^{2\pi}\int_0^\infty\textbf{R}_\textbf{n}\left(g\sqrt{\frac{m}{\bar{m}}}\right)p(\theta)q(m)\text{d}m\text{d}\theta\right)}_{:=\textbf{G}_{\bar{m};g}}\textbf{s}\cdot\boldsymbol{\sigma},
\end{equation*}
where $\textbf{n}=[\cos\theta,{\sin\theta},0]^\intercal$ and $p(\theta)$, $q(m)$ are the probability distributions followed by the phase and the photon number, respectively. The rotation matrix can be found through Rodrigues' formula: in general,
\begin{equation*}
\textbf{R}_\textbf{n}(\alpha)=\textbf{I}_3+(\sin\alpha)\textbf{N}+(1-\cos\alpha)\textbf{N}^2,\quad \textbf{N}=\begin{bmatrix}
        0 &-n_z&n_y\\
        n_z & 0 & -n_x\\
        -n_y & n_x & 0
    \end{bmatrix}.
\end{equation*}
To assist with analytical derivation, we assume that the average photon number $\bar{m}$ is large enough ($\gtrsim$100) such that $p(\theta)$ and $q(m)$ are well-approximated by normal distributions:
\begin{equation*}
q(m)=\frac{1}{\sqrt{2\pi\sigma_m^2}}e^{-\frac{1}{2}\left(\frac{m-\bar{m}}{\sigma_m}\right)^2},\quad p(\theta)=\frac{1}{\sqrt{2\pi\sigma_\theta^2}}e^{-\frac{1}{2}\left(\frac{\theta}{\sigma_\theta}\right)^2},
\end{equation*}
$\sigma_m$ and $\sigma_\theta$ being the corresponding variances. With these the integral can be computed to be
\begin{equation}\label{BlochMatrix}
\textbf{G}_{\bar{m};g}=\textbf{I}_3+\eta^{\frac{1}{4}}B\begin{bmatrix}
       0&0&0\\
       0&0&-1\\
       0&1&0
    \end{bmatrix}
    +\left(1-A\right)
    \begin{bmatrix}
       -\frac{1-\eta}{2}&0&0\\
       0&-\frac{1+\eta}{2}&0\\
       0&0&-1
    \end{bmatrix},\quad\eta=e^{-2\sigma_\theta^2},
\end{equation}
where
\begin{equation*}
A=\frac{\bar{m}}{\sqrt{2\pi\sigma_m^2}}\int_{-1}^\infty \cos\left (g\sqrt{1+t}\right)e^{-\frac{1}{2}\left(\frac{\bar{m}}{\sigma_m}\right)^2t^2}\text{d}t,\quad B=\frac{\bar{m}}{\sqrt{2\pi\sigma_m^2}}\int_{-1}^\infty \sin\left (g\sqrt{1+t}\right)e^{-\frac{1}{2}\left(\frac{\bar{m}}{\sigma_m}\right)^2t^2}\text{d}t.
\end{equation*}
In our case, the Bloch vector of the initial state, $\ketbra{0}{0}$, is $\textbf{s}_0=[0,0,1]^\intercal$. Since $\textbf{G}_{\bar{m};g}$ acts irreducibly on the $yz$ plane, we may restrict dynamics to this subspace. The restricted Bloch vector after each step evolves as
\begin{align}\label{rotatedBloch}
\begin{split}
\textbf{s}_{N}&=(\textbf{G}_{\bar{m};g})^N\textbf{s}_0\\
              &= \begin{bmatrix}
                  1-\frac{1}{2}(1-A)(1+\eta) & -\eta^{\frac{1}{4}}B\\
                  \eta^{\frac{1}{4}}B&A
              \end{bmatrix}^N
              \begin{bmatrix}
                  0\\1
              \end{bmatrix}\\
              &=\frac{r^{N-1}}{\sin\alpha}\begin{bmatrix}
                  -\eta^{\frac{1}{4}}B\sin(N\alpha)\\
                  A\sin(N\alpha)-r\sin((N-1)\alpha)
              \end{bmatrix},
\end{split}
\end{align}
where the two eigenvalues of $\textbf{G}_{\bar{m};g}$ are expressed as
\begin{align*}
re^{\pm i\alpha}=\frac{1}{4}\left(1-\eta+A(\eta+3)\pm i\sqrt{16\eta^{\frac{1}{2}}B^2-(1-A)^2(1-\eta)^2}\right).
\end{align*}
The state after the $N^{\text{th}}$ step  is $\rho_N=\frac12(\textbf{I}_2+\textbf{s}_N\cdot\boldsymbol{\sigma})$. From its Bloch representation, the QFI after the $N^{\text{th}}$ step with respect to the parameter $g$ can be readily calculated (see, for example, Ref.~\cite{Zhong2013}):
\begin{equation}\label{QFIexact}
F_N=\begin{cases}
|\partial_g\textbf{s}_{N}|^2+\frac{(\textbf{s}_{N}\cdot\partial_g\textbf{s}_{N})^2}{1-|\textbf{s}_{N}|^2}, & |\textbf{s}_{N}|<1;\\
|\partial_g\textbf{s}_{N}|^2, & |\textbf{s}_{N}|=1.
\end{cases}
\end{equation}
To continue the computation, we assume further that the light field has photon statistics corresponding to either a Poisson (coherent) or sub-Poisson distribution, such that $\sigma_m=k_m\sqrt{\bar{m}},\sigma_\theta=\frac{k_\theta}{2\sqrt{\bar{m}}}$ with $k_m\sim\mathcal{O}(1)$. Then, for $\bar{m} \gg g^2$, various terms may be approximated as
\begin{align*}
A&\approx \sqrt\frac{\bar{m}}{2\pi k_m^2}\int_{-\infty}^\infty \cos\left (g\left(1+\frac{t}{2}\right)\right)e^{-\frac{\bar{m}t^2}{2k_m^2}}\text{d}t=e^{-\frac{(gk_m)^2}{8\bar{m}}}\cos(g)\approx\left(1-\frac{(gk_m)^2}{8\bar{m}}\right)\cos(g);\\
B&\approx\left(1-\frac{(gk_m)^2}{8\bar{m}}\right)\sin(g);\quad \eta\approx1-\frac{k_\theta^2}{2\bar{m}},
\end{align*}
leading to
\begin{equation*}
\alpha\approx g,\, r\approx 1-\frac{\Delta(g)}{2\bar{m}},\quad \Delta(g)=\frac{k_m^2g^2+k_\theta^2(1-\cos(g))}{4}.
\end{equation*}
Similarly, the Bloch vector \eqref{rotatedBloch} and its derivative are
\begin{equation}\label{Blochapprox}
\textbf{s}_N = r^{N}\left(\begin{bmatrix}
    0\\ {-}\sin(Ng) \\ \cos(Ng)
\end{bmatrix}+\mathcal{O}\left(\frac{g^2}{\bar{m}}\right)\right),\quad \partial_g\textbf{s}_N = r^{N}\left({-}N\begin{bmatrix}
    0\\ \cos(Ng) \\ \sin(Ng)
\end{bmatrix}+\mathcal{O}\left(\frac{g}{\bar{m}}\right)\right).
\end{equation}
\begin{figure}[h!]
\includegraphics[width=0.5\textwidth]{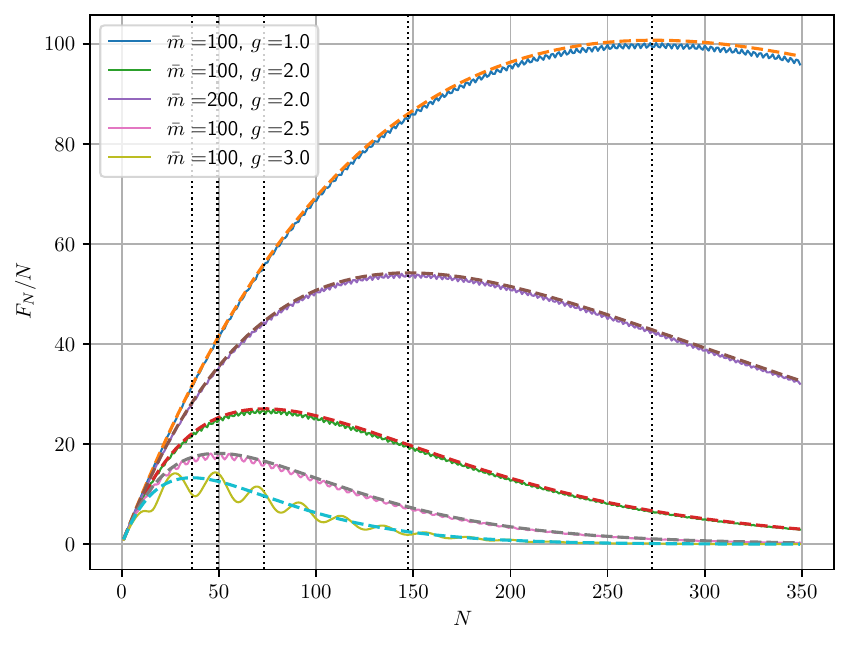}
\caption{Plots of $F_N/N$ for various $\bar{m}$ and $g$, when the field is in coherent states. Solid lines are the exact results using Eqs.~\eqref{rotatedBloch}, \eqref{QFIexact} and the dashed ones are the approximated ones \eqref{approxAppendix}. The vertical dotted lines approximate the optimal step $N_{\textrm{opt}}=-[2\log(r)]^{-1}$. As expected, the smaller and the larger $g$ and $\bar{m}$ are, respectively, the more negligible the percentage error becomes.}\label{plottQFI}
\end{figure}

The leading order effect of $\textbf{G}_{\bar{m};g}$ on the $yz$-plane is a rotation of angle $g$ combined with a shrinking of factor $r$. This coincides with the phase-covariant channel \eqref{covariant} by identifying $r$ with $\lambda_\perp$ (and replacing the $Z$-basis with $X$-basis). Eq.~\eqref{QFIapproxNichol} may thus be adopted to estimate the QFI, leading to
\begin{equation}\label{approxAppendix}
F_N(\bar{m};g)= N^2 r^{2N}\left(1+\mathcal{O}\left(\frac{g}{\bar{m}}\right)\right)\approx N^2\left(1-\frac{\Delta(g)}{2\bar{m}}\right)^{2N}.
\end{equation}
The merit of the approximation is confirmed by Figure~\ref{plottQFI} for coherent states. 

Finally, the number-phase uncertainty relation states that $\sigma_m\sigma_\theta\geq\frac{1}{2}$ and so $k_mk_\theta\geq1$ \cite{Pegg1989}. This implies the lower bound,
\begin{equation}\label{squeezed}
    \Delta(g)\geq\frac{k_mk_\theta g\sqrt{1-\cos(g)}}{2}\geq\frac{g\sqrt{1-\cos(g)}}{2},
\end{equation}
with equality attained at $k_m=g^{-1/2}(1-\cos(g))^{1/4}=k_\theta^{-1}$. A squeezed coherent state with real squeezing parameter $s$ has $k_m=e^{-s}$, $k_\theta=e^s$ and so the QFI is maximised at $s=\frac{1}{2}\log{\left(\frac{g}{\sqrt{1-\cos(g)}}\right)}$; notice this optimal squeezing level is dependent on the unknown parameter $g$. That said, in the main text we will set the control field to be in coherent states ($s=0$ and $k_m=k_\theta=1$ \cite{Gerry_Knight_2004}) due to their near-classical properties and easier experimental preparation.

\section{{Further Remarks on Energy Cost}}\label{sec:Landauer}
\subsection{{Cost of Discarding Subsystems}}\label{sec:LandauerReal}
{In Section~\ref{sec:energynotion} we set the energy cost $E$ to originate from energy conservation, and for each gate applied to the system qubit in the QPE protocol this cost amounts to $\bar m\hbar\omega\approx\bar m\hbar\omega_0$. Meanwhile, Landauer's principle~\cite{Landu,LanduLloyd_1997} states that the minimal energy cost of erasing information from one qubit is on the scale of $k_BT_0$, with $T_0$ being the system temperature. The relative magnitude between the two fundamental energy costs is then
\begin{equation}\label{ratioEng}
\frac{E_{\text{conservation}}}{E_{\text{Landauer}}}\sim \frac{\bar m}{\xi},\quad    \xi=\frac{k_BT_0}{\hbar\omega_0}.
\end{equation}
Using the value $\xi \approx 0.2$ as quoted in Section~\ref{sec:stateprep} under typical experimental settings, one sees that, for the range of photon number $\bar m\gtrsim 100$ considered in our semi-classical approximation on the EM field, the energy cost from conservation as defined in Eq.~(\ref{def:energy}) is at least several orders of magnitude larger than the fundamental thermodynamic erasure cost arising from Landauer's principle, justifying the omission of the latter in our analysis. This in particular validates our assumption in Table~\ref{tableenergy} that discarding ancillary qubits can be regarded as a free operation. As per the fundamental thermodynamic cost of discarding the field state after implementing each phase-encoding channel, note that the back-action on the field due to the field-qubit interaction~\eqref{jaynes} is small in our semi-classical regime, meaning that the post-evolution field state is still very close to a pure state (with vanishing temperature), and its discarding cost is consequently assumed to be negligible as well compared to the total energy of the field. In the following subsection, we briefly discuss how  reusing or recycling the post-evolution field state may incur further energy savings.}

\subsection{{Energy Saving Strategies}}\label{sec:recycle}
{Here we qualitatively describe two possible generalisations to reduce the total energy cost, both of common practice. First, as shown in Appendix~\ref{appendix:computingQFI}, $\Delta(g)$ may be reduced according to Eq.~\eqref{squeezed}, if squeezing is introduced to the EM field. For a fixed $\bar m$ this leads to a smaller complexity (see Eq.~\eqref{approxQFIComp}), at the cost of extra squeezing energy per gate (the energy of a squeezed coherent state with real squeezing is $\bar m+\sinh^2(s)$). An energy-optimal squeezing level can be accordingly determined. 

The energy cost can be further saved through field recycling. After the implementation of each channel, the field state is slightly entangled with the system qubit and its final (mixed) state will marginally deviate from the initial (pure) state, leading to larger implementation error for the next iteration if we reuse the final field state as it is. To circumvent such back-action --- which, albeit small in each gate implementation, can accumulate significantly throughout the full QPE protocol --- in the main text we completely reset the field after each phase-encoding gate is implemented. Note that this may also be the result of an operational constraint, where each query to the agent in charge of implementing the gate must be made independently and hence the implementation error at each step is uncorrelated from the previous history. Instead, we can imagine that some restoring procedure may be allowed to recycle the same field for the next step, hence saving the reinitialisation energy of the field. For example, as studied in Ref.~\cite{Ikonen_2017,Vuglar2024}, consider a squeezed coherent state after interacting with the system qubit and slightly perturbed by the back-action. The original field state can be well restored by sequentially interacting the field with an additional ancilla qubit for some carefully tailored duration, through the free interaction~\eqref{jaynes}. The more times the interaction is repeated, the closer the final field state will be to the original squeezed coherent state, and the smaller the implementation error will be for the next step, at the cost of larger ancilla overhead. This leads to another complexity versus (recycling) energy trade-off relation. The improvement may also be enhanced if monitoring is allowed on the field~\cite{Catana2014,Albarelli_2018}. A precise treatment on the energy reduction brought by more sophisticated field manipulation procedures is left to a later work.}


\subsection{{Error Mitigation Controls}}\label{sec:control}
{Although we aim to keep the current QPE protocol simple to demonstrate the qualitative behaviour of its energy cost from basic building blocks, it is interesting to investigate whether (in addition to varying the number of photons spent per gate) adding intermediate control operations on the probe system, as illustrated in Figure~\ref{fig:placeholder}, may help improve the QPE performance. 

\begin{figure}[hbt]
    \centering
\begin{tikzpicture}
        \draw(0,0)--(0.5,0);
        \draw (0.5,-0.5) rectangle (1.5,0.5)node[anchor=center,midway]{$\mathcal{G}_{\bar{m};g}$};
        \draw(1.5,0)--(1.75,0);
        \draw (1.75,-1) rectangle (2.75,0.5)node[anchor=center,midway]{$C_1$};
        \draw(2.75,0)--(3,0);
        \draw (3,-0.5) rectangle (4,0.5)node[anchor=center,midway]{$\mathcal{G}_{\bar{m};g}$};
        \draw(4,0)--(4.25,0);
        \draw (4.25,-1) rectangle (5.25,0.5)node[anchor=center,midway]{$C_2$};
        \draw(5.25,0)--(5.5,0);
        \draw(0,-0.75)--(1.75,-0.75);
        \draw(2.75,-0.75)--(4.25,-0.75);
        \draw(5.25,-0.75)--(5.5,-0.75);
         \filldraw (5.6,-0.375) circle (0.5pt);
         \filldraw (5.7,-0.375) circle (0.5pt);
         \filldraw (5.8,-0.375) circle (0.5pt);
        
    \end{tikzpicture}
    \caption{{A typical ancilla-assisted control sequence used to suppress the noise on the system qubit.}}
    \label{fig:placeholder}
\end{figure}

Recall the implemented channel, whose action on the Bloch sphere is described by Eq.~\eqref{Blochapprox}, can be approximated as the desired unitary operation $U_g$ followed by a bit flip channel with flipping probability $p=\frac{1-r}{2}=\frac{\Delta(g)}{4\bar m}$:
\begin{equation*}
\mathcal{G}_{\bar{m};g}(\rho)=(1-p)U_g\rho U_g^\dag+p\sigma_xU_g\rho U_g^\dag \sigma_x.
\end{equation*}
The goal of the control operations would then be to reduce the bit flip noise.

However, since in our case the noise generator $\sigma_x$ coincides with the Hamiltonian that generates the signal, it is fundamentally impossible to isolate and suppress the noise. Indeed, results in Ref.~\cite{RafalSeq} indicate that the QFI obtained in Eq.~\eqref{QFIapproximated} is optimal within our sequential framework. Common error mitigation techniques, such as quantum error correction~\cite{Sisiancilla,Layden_2019,Zhou_2018,Lukin2014}, feedback controls~\cite{RafalQEC,Sekatski_2017}, and passive ancillae~\cite{fujiwara_imai_2008,Demkowicz_Dobrza_ski_2012,Nichols2016} cannot help increase the QFI; in other words,  for our protocol the best possible control operations $C_1, C_2, \ldots$ in Figure~\ref{fig:placeholder} all reduce to the identity. Further improvement is only possible if we introduce entanglement amongst probes of different sequences. We refrain from doing so since combining energy and entanglement, as two distinct resource quantities, demand a more careful bookkeeping on the total resource cost, a point brought up again in the concluding Section~\ref{section:outlook}. 

The aforementioned error mitigation techniques nevertheless becomes effective when other types of noise take place --- such as decoherence arising from system-environment interaction that does not commute with $U_g$; common examples include dephasing and depolarisation~\cite{Nielsen_Chuang_2010}. If the field state is recyclable as described in Section~\ref{sec:recycle}, the implementation errors at each step are correlated through the shared field (as well as other potential sources of memory effects from the environment) and the effective noise becomes non-Markovian, which can be suppressed e.g.~through dynamical decoupling procedures~\cite{Viola_1998,Alvarez2011}. Even within the current scope, when the photon number $\bar m$ is small enough, apart from the leading order bit flip noise, non-commuting terms also start to have a significant effect. Intuitively, we anticipate the appearance of yet another complexity versus (control) energy trade-off relation, where the error mitigation control increases the total QFI and hence reduce the complexity, while introducing extra energy cost for its implementation. A more thorough analysis on such an interplay will be explored in a forthcoming work.}

\section{{Finite-Sample Effects on Estimator Variance}}\label{app:variance}

{
The sweet spot discovered in Section~\ref{EngtoComp} corresponds to the regime where a single sequence suffices to reach the target total QFI, and hence the desired variance bound $\delta^2$ implied by the quantum Cram\'er--Rao bound (QCRB)~\eqref{cramer}. While the QFI provides a well-established figure of merit, leading to a convenient and computable constraint when determining the complexity in Eq.~\eqref{def:complexity}, it does not automatically reflect the achievable estimation precision in the finite-sample regime. This is because, as remarked in the main text, the QCRB assumes an unbiased estimator from the optimal measurement, which in general only holds asymptotically when the sequence is repeated for a sufficiently large number of times, $Q\gg1$, and adaptive strategies are implemented~\cite{Giovannetti_2004,Giovannetti2006,Giovannetti_2011}. 

To assess the impact of finite-sample effects, one may consider alternative bounds on the estimator variance that do not rely on asymptotic or unbiasedness assumptions, such as the  Bayesian generalisation of the Cram\'er--Rao bound based on the van Trees inequality~\cite{Trees}. In what follows, we focus on the quantum Ziv--Zakai bound (QZZB) as a universally valid lower bound that holds for any estimator bias~\cite{Tsang_2012} and is therefore applicable even in the single-shot regime $Q \sim 1$.
Assuming a uniform prior distribution for $g$ over an interval of width $\Gamma$ and mean $\mu$, the QZZB reads
\begin{equation}\label{eq:QZZB}
\text{Var}[\hat g]\geq  \frac{1}{2\Gamma}\int_0^\Gamma\text{d}x\,x\int_{\mu-\frac{\Gamma}{2}}^{\mu+\frac{\Gamma}{2}-x}\text{d}g \left[1-\sqrt{1-F\big(\rho_N(g),\rho_N(g+x)\big)}\right],
\end{equation}
where $F(\cdot,\cdot)$ is the quantum fidelity between two states and $\rho_N(g)$ is the final state after $N$ steps, for a value $g$ of the parameter to be estimated. 
For a qubit system, the fidelity between two states with Bloch vectors $\textbf{s}_1$ and $\textbf{s}_2$ is
\begin{equation*}
F(\rho_1,\rho_2)=\frac{1}{2}\left(1+\textbf{s}_1\cdot\textbf{s}_2+\sqrt{(1-|\textbf{s}_1|^2)(1-|\textbf{s}_2|^2)}\right).
\end{equation*}
In our setting, using Eq.~\eqref{Blochapprox} for the Bloch vector and working in our usual semi-classical regime specified by $\bar{m} \gtrsim 10^2$, we can approximate the shrinking factor with the constant $r\approx 1-\frac{\Delta(\mu)}{2\bar m}$. Under this approximation, the fidelity  becomes independent of $g$ and reads
\begin{equation*}
F\big(\rho_N(g),\rho_N(g+x)\big)\approx 1-r^{2N}\sin^2\left(\frac{Nx}{2}\right).
\end{equation*}
Substituting into the QZZB (\ref{eq:QZZB}) we get
\begin{equation*}
\text{Var}[\hat g]\gtrsim \frac{1}{2}\int_0^\Gamma\text{d}x\,x\left(1-\frac{x}{\Gamma}\right)\left(1-r^N|\sin(Nx/2)|\right).
\end{equation*}

For large $N\approx N_{\textrm{opt}}$, as done in  Eq.~\eqref{optimalstepapprox}, we use the approximation $r^N\approx e^{-1/2}$ and replace the oscillatory term by its average value, $|\sin(Nx/2)|\to 2/\pi$, obtaining
\begin{equation}\label{eq:QZZBapprox}
\text{Var}[\hat g]\gtrsim k\Gamma^2,\quad k=\frac{1-\frac{2}{\pi\sqrt{e}}}{12}\approx0.05.
\end{equation}

This bound provides a finite-sample benchmark that can be compared with the asymptotic QCRB. For the QFI derived in the main text to reflect the achievable estimation precision, we require the QCRB \eqref{cramer} to be at least as informative as the QZZB bound \eqref{eq:QZZBapprox}.
This is satisfied if, again using the derivation in Eq.~\eqref{optimalstepapprox},
\begin{equation*}
\frac{1}{F_{N_{\textrm{opt}}}}\approx e\left(\frac{\Delta(\mu)}{\bar m}\right)^2\gtrsim k\Gamma^2\,\Rightarrow\,\Gamma\lesssim \frac{14.8}{\bar m},
\end{equation*}
for $\mu\approx2.5$ as adopted in Figures~\ref{plot:QPEEnergy},~\ref{coolingexample},~\ref{figure:mmtcost}. 
Therefore, a smaller $\bar m$ allows for a less stringent demand on the prior knowledge on $g$, translating into a more informative QCRB in the finite-sample regime. Furthermore, recall that the constraint to determine the complexity~\eqref{QPEcomplexity} is a desired lower bound $\delta^2$ on the estimator variance. Naturally, this bound should be at least tighter than the one given by the prior uniform distribution on $g$, meaning that $\delta^2\lesssim \Gamma^2/12$, and thus
\begin{equation*}
    \bar m\lesssim\frac{4.3}{\delta}.
\end{equation*}
For our choice of $\delta^2=10^{-4}$ in the main text, this imposes a constraint on the photon number, $\bar m\lesssim 5\cdot10^2$. 

Combining the above considerations, we identify a {\em Goldilocks regime}
\begin{equation}\label{sweeter}
10^2 \lesssim \bar m \lesssim 5\cdot10^2,
\end{equation}
in which the assumptions underlying the analysis in the main text --- most notably the semi-classical approximation --- are satisfied and the QCRB remains an informative benchmark for the achievable estimation precision despite finite-sample effects. Consequently, the sweet spot identified in Section~\ref{EngtoComp} yields a physically meaningful resource trade-off.

Overall, the analysis in this Section indicates that operating at smaller photon number per gate not only reduces the implementation energy, but also improves the consistency between finite-sample performance and the asymptotic precision benchmark.
}

\section{Classical Fisher Information from Imperfect Measurements}\label{appendix:mmt}
Following from the pointer model introduced at the beginning of Section~\ref{sec:mmt}, suppose {the optimal measurement that yields the QFI is represented by the POVM operators $\textbf{M}=(M_1,M_2)$ and we perform it on the pointer state.} If the pointer is initially in the pure state $\ketbra{0}{0}$, then it can be checked that the corresponding POVM operators on the system qubit is exactly $\textbf{M}$ as desired. However, instead the pointer is prepared in a thermal state \eqref{thermalstate} after a dynamic cooling procedure as described in Section~\ref{sec:stateprep}. For simplicity we assume the cooling qubits are the same as those during the state preparation stage, although in the main text they may have different initial temperature and transition frequency depending on experimental realisation. The same measurement procedure now yields the modified POVM operators,
\begin{equation*}\label{noisyPOVM}
    \tilde{\textbf{M}}=(1-\varepsilon)\textbf{M}+\varepsilon\textbf{N},{\quad\textbf{N}=(\sigma_x M_1\sigma_x, \,\sigma_x M_2\sigma_x)},\quad\varepsilon=\frac{1-\gamma(T)}{2},
\end{equation*} 
where $T\approx 2T_0/M_m$ and $M_m$ is the number of qubits consumed to cool the pointer. Consequently, the estimation precision is reflected by the CFI \eqref{CFI} achieved by the non-optimal $\tilde{\textbf{M}}$, rather than the QFI achieved by $\textbf{M}$. To compute the effect of this imperfect measurement, we adopt the results from Ref.~\cite{Kurdzia_ek_2023}. The susceptibility of the CFI with respect to a small disturbance on the POVM operators is defined as
\begin{equation*}
\chi[\textbf{M},\textbf{N}]=\lim_{\varepsilon\rightarrow0}\frac{F_c[\textbf{M}]-F_c[\tilde{\textbf{M}}]}{\varepsilon F_c[\textbf{M}]}.
\end{equation*}
In our case, $F_c[\textbf{M}]=F_{N}(\bar{m};g)$. The CFI can be approximated up to the leading order of the perturbation $\varepsilon$ as $F_c[\textbf{M}]\left(1-\varepsilon\chi[\textbf{M},\textbf{N}]\right)$. By plugging in Eq.~\eqref{CFI} explicitly and using {$M_1+M_2=N_1+N_2=I_2$, $\Tr(\rho_N)=1$, $\Tr(\dot\rho_N)=0$}, this can be simplified to
\begin{equation*}
F_{N}(\bar{m};g)\xrightarrow{\text{measurement}} F_{N}(\bar{m};g,\varepsilon)=F_{N}(\bar{m};g)-{\left(\frac{\dot p^2(q-p)(1-2p)}{p^2(1-p)^2}-\frac{2\dot p(\dot q-\dot p)}{p(1-p)}\right)}\varepsilon+\mathcal{O}(\varepsilon^2),
\end{equation*}
{where $p=\Tr(M_1\rho_N)$, $q=\Tr(N_1\rho_N)=\Tr(\sigma_xM_1\sigma_x\rho_N)$ and $\dot p=\Tr(M_1\partial_g\rho_N)$, $\dot q=\Tr(N_1\partial_g\rho_N)$. To determine the optimal POVM}, recall that while extracting the leading order contribution to the QFI \eqref{approxAppendix}, the Bloch vector of the probe state is approximated by Eq.~\eqref{Blochapprox}; in particular, $\textbf{s}_N\cdot\partial_g\textbf{s}_N\approx 0$. The SLD operator can then be easily guessed as $\Lambda_g=\partial_g\textbf{s}_N\cdot\boldsymbol{\sigma}=2\partial_g\rho_N$ (also see Ref.~\cite{Zhong2013}). The POVM operators that maximise the CFI are projectors onto the eigenspaces of $\Lambda_g$,
\begin{equation}\label{eqn:optimalPOVM}
M_i=\frac{\textbf{I}_2+\hat{m}_i\cdot\boldsymbol{\sigma}}{2},\quad\hat{m}_i\approx\pm\frac{\partial_g\textbf{s}_N}{|\partial_g\textbf{s}_N|},
\end{equation}
leading to
{
\begin{align*}
p&=\frac{1+\hat m_1\cdot \textbf{s}_N}{2}\approx\frac{1}{2},\quad q=\frac{1-\hat m_i\cdot \textbf{s}_N}{2}\approx\frac{1}{2};\\
\dot p&=-\dot q=\frac{\hat m_1\cdot\partial_g \textbf{s}_N}{2}\approx\frac{|\partial_g\textbf{s}_N|}{2},
\end{align*}
where we use the fact that $\textbf{s}_N$ always stays in the $yz$-plane and so a $\sigma_x$-rotation merely flips its sign. Therefore, the leading order CFI is
\begin{equation*}
F_N(\bar m;g,\varepsilon)\approx F_N(\bar m;g)-4|\partial_g\textbf{s}_N|^2\varepsilon\approx(1-4\varepsilon)F_{N}(\bar{m};g),
\end{equation*}}
where the last relation comes from Eq.~\eqref{QFIexact}.

Combined with Eq.~\eqref{coolscaling}, the overall modification to the QFI due to non-ideal state preparation and measurement is
\begin{equation}
F_{N}(\bar{m};g)\xrightarrow{\text{state preparation + measurement}}F_N(\bar{m};g,M_s,M_m)\approx\left[\gamma\left(\frac{2T_0}{M_s}\right)\right]^2\left[2\gamma\left(\frac{2T_0}{M_m}\right)-1\right]F_{N}(\bar{m};g).
\end{equation}
The resulting complexity and total energy cost can be derived in the same fashion as in Section~\ref{sec:stateprep}.
\clearpage


\end{document}